\documentclass[reprint,floatfix,prb,aps]{revtex4-2}
\pdfoutput=1

\usepackage{graphicx}
\usepackage{dcolumn}

\newcommand\numberthis{\addtocounter{equation}{1}\tag{\theequation}}
\newcommand{\rom}[1]{\text{\uppercase\expandafter{\romannumeral #1\relax}}}
\usepackage{lineno}
\usepackage[toc,page]{appendix}

\usepackage{amsmath}
\usepackage{fancyhdr}
\usepackage{amssymb}
\usepackage{amsthm}
\usepackage{dsfont}
\def\nn{\nonumber}
\newcommand{\eq}[1]{\begin{align*}#1\end{align*}}

\def\p{\par}
\def\mbf{\mathbf}

\usepackage{hhline}
\usepackage{diagbox}
\usepackage{multirow}

\newcommand*{\ket}[1]{|{#1} \rangle}

\newcommand*{\refig}[1]{Fig.~\ref{fig:#1}}

\usepackage[%
    colorlinks=true,
    pdfborder={0 0 0},
    linkcolor=red,
    citecolor=blue
]{hyperref}

\makeatletter
\newcommand{\customlabel}[2]{%
   \protected@write \@auxout {}{\string \newlabel {#1}{{#2}{\thepage}{#2}{#1}{}} }%
   \hypertarget{#1}{}
}
\makeatother

\begin{document}

\title{Phase transitions in $(2+1)$D subsystem-symmetric monitored quantum circuits} 
\author{Cole Kelson-Packer}
\email{ckelsonpacker99@unm.edu}
\affiliation{Center for Quantum Information and Control, Department of Physics and Astronomy, University of New Mexico, Albuquerque, New Mexico 87106, USA}
\author{Akimasa Miyake}
\email{amiyake@unm.edu}
\affiliation{Center for Quantum Information and Control, Department of Physics and Astronomy, University of New Mexico, Albuquerque, New Mexico 87106, USA}
\date{\today}
\begin{abstract}
    \p The interplay of unitary evolution and projective measurements is a modern interest in the study of many-body entanglement. On the one hand, the competition between these two processes leads to the recently discovered measurement-induced phase transition (MIPT). On the other hand, measurement-based quantum computation (MBQC) is a well-known computational paradigm where measurements simulate unitary evolution by utilizing the entanglement of special resources such as the two-dimensional (2D) cluster state. The entanglement properties enabling MBQC may be attributed to symmetry-protected topological (SPT) orders, particularly subsystem-symmetric topological (SSPT) orders. It was recently found that the one-dimensional cluster state may be associated with an SPT phase in random circuits respecting a global $\mathbb{Z}_2\times\mathbb{Z}_2$ symmetry, and furthermore that all phase transitions in this scenario belong to the same universality class. As resources with greater computational power feature greater symmetry, it is fruitful to investigate further any relationship between levels of symmetry in MIPTs and MBQC. In this paper we investigate MIPTs on a torus with three levels of symmetry-respecting unitary evolution interspersed by measurements. Although we find two area-law phases and one volume-law phase with distinct entanglement structures for each ensemble, the phase transition from the volume-law phase to the area-law phase associated with the 2D SSPT cluster state has variable correlation length exponent $\nu$. Whereas $\nu\approx 0.90$ for unconstrained Clifford unitaries and $\nu\approx0.83$ for globally-symmetric Cliffords, subsystem-symmetric Cliffords feature a much smaller value $\nu\approx 0.38$. We discuss how these distinct $\nu$'s quantify spacetime response scales where quantum information is manipulated by single-qubit measurements as in MBQC.
\end{abstract}
\maketitle

\section{Introduction}
\label{sec:intro}
\p Quantum computers, besides offering prospective advantages in certain computational tasks \cite{Fey82,Shor97}, and the ability to realize classically impossible effects \cite{Bell64,Cabello01,Cabello01p}, also provide tunable platforms for investigating various many-body phenomena. It is well-established that random monitored circuits, wherein both unitary operations and mid-circuit measurements are carried out on a set of qubits, provide novel insights into the dynamics of quantum entanglement via the measurement-induced phase transition (MIPT). The phenomenology of this transition has been the subject of many recent works \cite{LiChenFisher18,ChenFisher19,SkinnerRuhmanNahum19,GullansHuse20,ChoiBaoQiAltman20,ChoiBaoQiAltman20p,VasseurFisherLudwig21}, and several reviews are available \cite{PotVass21,Skinner23,Fish23}. 
\p The central fact is that the competing local processes of entanglement-producing unitaries and disentangling measurements are balanced at a nonzero measurement rate. Along one line of intuition, this is surprising; for two subsystems of a pure state, only unitaries across their boundary produce entanglement between them, whereas measurements may have a deleterious effect anywhere. According to this perspective any finite projection rate should destroy entanglement in the thermodynamic limit \cite{ChanNandkishorePretkoSmith19}. While this logic follows through in certain systems \cite{CaoTilloryDeLuca19}, remarkably more often than not a stable ``volume-law'' entanglement phase exists in addition to a low entanglement ``area-law'' counterpart \cite{LiChenFisher18,ChenFisher19,SkinnerRuhmanNahum19}.
\p The importance of these random monitored circuits (and the MIPT more generally) is that their phenomenology is thought to reflect behavior of generic monitored systems, just as random matrix theory qualitatively describes heavy atomic nuclei \cite{Skinner23,Wigner55}. More specialized scenarios may be investigated by tuning the setup. One such generalization introduces additional stabilizer measurements. Taking these to be stabilizers of states of particular interest allows one to probe general entanglement properties of the classes such states belong to. For example, $(2+1)$D random circuits with toric code measurements feature a phase diagram with a stable critical region due to the robustness of the associated stabilizer state \cite{LavasaniAlaviradBarkeshli21p}. Similarly, a symmetry-protected topological (SPT) phase has been realized in $(1+1)$D by considering stabilizers of the one-dimensional (1D) cluster state and restricting the unitaries to respect a global $\mathbb{Z}_2\times\mathbb{Z}_2$ symmetry \cite{LavasaniAlaviradBarkeshli21}. 
\p One result of this last study is that the critical exponents of the transition to the SPT phase match $(1+1)$D percolation, with correlation length exponent $\nu\approx 4/3$ in particular. Although diagnostics like the topological entanglement entropy differentiate the SPT phase from the trivial phase, all transitions in the phase diagram belong to the same universality class. This is true regardless of whether the transition moves from the volume-law phase to the SPT phase, or to the trivial area-law entanglement phase. 
\p In this sense the transition would seem insensitive to the special properties of the 1D cluster state. One such property is the fact that the 1D cluster state is a computational resource for measurement-based quantum computing (MBQC). MBQC is an alternative to the traditional gate model based on the measurement protocols carried out on special resource states \cite{BriegelRauss01,RaussBriegel01,RaussBrowneBriegel03}. While the 1D cluster state is not a resource for universal computation, it acts as a quantum wire, teleporting edge-encoded states by adaptive measurements. Most quantum states cannot be used for even this purpose. Rather, it is a consequence of the special entanglement structure of the 1D cluster state emerging from $\mathbb{Z}_2\times\mathbb{Z}_2$ symmetry \cite{DohertyBartlett09,ElseSchwarzBartlettDoherty12}. For universal computation the requisite symmetry structure is even more particular; not only do states with little entanglement lack the necessary correlations to achieve this effect, but even highly entangled states fail to be useful \cite{GrossFlammiaEisert09,BremnerMoraWinter09}. It is an intriguing question to what extent this hierarchy of symmetry/utility in MBQC has a counterpart in MIPT---that is, how the transitions are affected by symmetry.
\p To this end we consider $(2+1)$D random monitored quantum circuits featuring---in addition to the usual single-qubit measurements---two-dimensional (2D) cluster state stabilizer measurements. We analyze three different ensembles of unitary dynamics. As defined later, these unitary ensembles consist of five-qubit local Cliffords with no constraints (``Clifford''), global symmetry-protected topological constraints (``SPT Clifford''), and subsystem symmetry-protected topological constraints (``SSPT Clifford''). These ensembles are associated with a hierarchy of symmetries of the 2D $L\times L$ square lattice, $\mathds{I}\subset \mathbb{Z}_2\times\mathbb{Z}_2\subset \mathbb{Z}_2^{2L-1}$. Using entanglement diagnostics to detect the transition, we find that the critical exponents of the transition---and thus the associated universality classes---differ between these unitary ensembles.
\p In particular, we find that while the Clifford ensemble features transitions between both area-law phases and the volume-law phase with the same critical exponent as $(2+1)$D percolation, the SSPT Clifford ensemble exhibits new exponents differing from both each other and percolation. This suggests that our SSPT circuits---respecting the subsystem symmetry associated with universal measurement-based quantum computation---produce states with more selective entanglement properties.
\p Following this Introduction, Sec.~\ref{sec:background} provides the background for our setting in Subsec.~\ref{subsec:setting}. Then, after adumbrating our numerical methods in Subsec.~\ref{subsec:num}, we deliver our results in Subsec.~\ref{subsec:results}. This is split into three further subsubsections for each unitary ensemble considered, with an additional fourth section investigating the suppression of the volume-law phase. We conclude with a discussion in Sec.~\ref{sec:discussion}. Our Appendix includes elaborations of various minutiae, including further details of our numerical methodology, elucidation of order parameters, and analysis of properties at the pure-measurement critical point.

\section{Background}
\label{sec:background}
\begin{figure}
    \centering
    \includegraphics[width=0.9\linewidth]{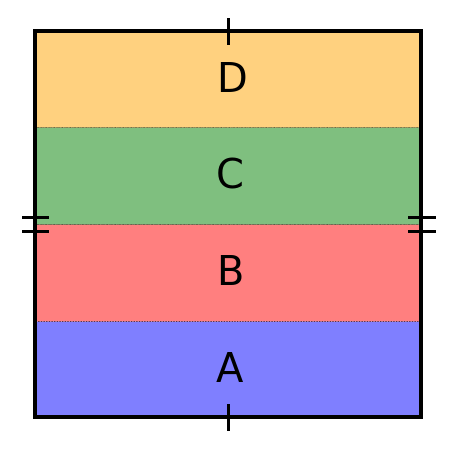}
    \caption{\label{fig:q1dregions} Our regions chosen for evaluating the topological entanglement entropy $S_{\text{top}}$ are cylinders on a torus. This can be represented as rectangles on a 2D plane with opposite sides identified.}
\end{figure}

\p Random monitored quantum circuits present a novel setting for investigating generic many-body quantum phenomena where both the effects of repeated measurements and unitary evolution are key players. When measurements are applied at some rate $p$, a so-called MIPT is observed between an area-law phase of low entanglement at high $p$ and a volume-law entanglement regime of high entanglement for low values \cite{LiChenFisher18,ChenFisher19,SkinnerRuhmanNahum19}. Measurements typically have a disentangling effect wherever they are applied, localizing or removing correlations. Unitary dynamics, meanwhile, spread information across long-range correlations. The transition between the volume- and area-law phases reflects whether or not ensembles of circuits generically ``protect'' quantum correlations encoded in the initial state \cite{ChoiBaoQiAltman20,ChoiBaoQiAltman20p}. Variations---such as additional measurement types or constraints on the unitaries applied---realizes further novel phases \cite{LavasaniAlaviradBarkeshli21,LavasaniAlaviradBarkeshli21p,SangHsieh21,BaoChoiAltman21,LiuZhouChen22,NegariSahuHsieh24}.
\p To understand the basic phenomenology of these circuits it is useful to map them to more standard mathematical models. For the case of $p=0$, where only pure unitary dynamics are in play, many of the observed entanglement properties can be related to the famous Kardar-Parisi-Zhang universality class \cite{NahumRuhmanVijahHaah17}. Entanglement growth has a parallel in ballistic deposition, operator spreading follows exclusion processes, and bounds on entanglement between regions is echoed by polymers in random energy environments. This last relationship is especially pertinent for nonzero $p$, as the problem changes into first passage percolation---another standard mathematical model with well-studied phase transitions \cite{Grim99}. In particular, the correlation length critical exponent of percolation in 2D, $\nu=4/3$, matches critical exponents found for the measurement induced phase transition in $(1+1)$D \cite{ChenFisher19,SkinnerRuhmanNahum19}. The intuition behind these mappings is supported by work exactly mapping certain select models to classical models with percolation exponents, such as the Potts model \cite{ChoiBaoQiAltman20,JianYouVasssLud20}.
\p There is good numerical agreement with these analytical results. It is perhaps surprising that in spite of the esoteric, nonequilibrium nature of this transition, many standard numerical tools may still be fruitfully applied. In particular, the finite-size scaling ansatz \cite{KawashimaIto93,KawashimaRieger97,HoudayrHartmann04} retains its utility. This workhorse utilizes the often-vindicated hypothesis that, near a transition point $p_c$, an order parameter $\Delta$ follows a certain function of the tuning parameter $p$ and system length $L$:

\eq{
    \Delta(L,p)\approx L^{\gamma}F((p-p_c)L^{1/\nu}), \numberthis\label{eq:fss}
}

where $F$ is some common function, and $\nu$ and $\gamma$ are universal exponents. By finding a good fit for numerical values of the diagnostic $\Delta$, $\nu$ and $\gamma$ may be estimated. The tuning parameter $p$ in this case refers to some measurement rate. As for the order parameter $\Delta$, an important fact to keep in mind is that the transition takes place on the level of nonlinear properties of individual quantum trajectories. For example, a transition emerges when the entanglement of individual states averaged over trajectories and circuits, $\overline{S_{vN}(\rho)}$, is considered, but disappears for the entanglement of the average state, $S_{vN}(\bar{\rho})$. This is because the entropy is nonlinear in the state---that is, because $\overline{S_{vN}(\rho)}\ne S_{vN}(\overline{\rho})$. Many useful diagnostics of the transition, such as the topological- or ancilla entanglement entropies \cite{KitaevPreskill06,LevinWen06,GullansHuse20p,GullansHuse20} are also nonlinear. This makes experimental realization of the transition's phenomenology highly challenging \cite{ion,Google23a,midmeas}.
\p ``The'' topological entanglement entropy in fact refers to a class of related metrics. The essential idea is this: from locality, the von Neumann entropy of a subsystem $A$ is expected to obey an area-law with subleading corrections, $S_A\sim |\partial A|-\Gamma$. Topological entanglement entropies $S_{\text{top}}$ are constructed by taking special combinations of subsystems and summing their entropies together in such a way as to isolate the subleading $\Gamma$. For topological quantum field theories this remainder is related to their quantum dimension, and more generally it is thought to reflect long-range entanglement properties. Taking sufficiently large system sizes washes out short range effects, yielding a robust measure of generic entanglement phenomena.
\p A common version is the seven-term topological entanglement entropy, also known as the tripartite mutual information \cite{KitaevPreskill06,ZengChenZhouWen19}. For subregions $A$, $B$, and $C$ of a system---such as for a torus, as depicted in \refig{q1dregions}---this may be defined as 

\small
\eq{
    S_{\text{top}}\equiv S_{AB}+S_{BC}+S_{AC}-S_{A}-S_B-S_C-S_{ABC}. \numberthis\label{eq:top}
}\normalsize

\p $S_{\text{top}}$ is particularly valuable when considering generalizations of the random monitored circuit setup involving special measurements and phases. If the measurements are of stabilizers of a state with a particular value of $S_{\text{top}}$, then the trajectory average of this value readily distinguishes an associated circuit phase. Taking the toric code $S_{\text{top}}$ as an example, the four-term variant of $S_{\text{top}}$ (related to strong subadditivity) and seven-term version related above yield means of $1$ or $2$, respectively. This differs from the $0$ of simple product states, and also from the extensive scaling of states generically produced by Haar unitary dynamics \cite{LavasaniAlaviradBarkeshli21p}.
\p Another guiding example is the 1D cluster state defined on an open chain \cite{LavasaniAlaviradBarkeshli21}. Where the toric code realizes a genuine topological phase, the 1D cluster state is a symmetry-protected topological (SPT) state. In 1D such states are classified by the projective representations of the eponymous symmetry group acting on the boundary---that is to say, by the second group cohomology \cite{ChenGuLiuWen13}. The associated phase in a random monitored quantum circuit also exhibits SPT behavior; $S_{\text{top}}$ here detects the edge degrees of freedom, again distinguishing the trivial phase from a more exotic SPT phase. The ancilla entanglement entropy also couples to these boundary degrees of freedom, reflecting the long-time storage of encoded edge qubits.
\p The cluster state behavior is particularly notable given the state's relationship with MBQC, a well-known alternative to the gate model of quantum computation. Instead of evolving a state through a circuit and enacting entangling operations before a final measurement, one begins with a special entangled ``resource state'' and simulates a circuit via evolution with adaptive local measurements. A particular class of resource states are the so-called ``graph states,'' which are states defined on a graph, with qubits prepared in the $\ket{+}$ state on vertices, and edges denoting qubits between which controlled-$Z$ operators are enacted. Equivalently, graph states can be said to be the joint eigenstate of the commuting operators $\sigma_x^\alpha\bigotimes\limits_{\beta\in N(\alpha)}\sigma_z^\beta$, where $\alpha,\beta$ label vertices and $N(\alpha)$ denote the graphical neighbors of $\alpha$. We will work with a slightly different convention, where a Hadamard is acted upon every qubit after this setup, exchanging the roles of the Pauli operators $X$ and $Z$.
\p The power of resource states is often attributable to underlying symmetries, whose presence reflect a kind of irreducible entanglement structure \cite{Miyake10}. Cluster states, defined on graphs of regular lattices, exhibit such symmetries. The 1D cluster state, for example, features $\mathbb{Z}_2\times\mathbb{Z}_2$ global symmetry, the factors flipping spins on even/odd sites. This symmetry enables the state to act as a 1D wire, facilitating the teleportation of an encoded edge qubit by measurement \cite{ElseSchwarzBartlettDoherty12}. Similarly, the 2D cluster state, enjoying an extensive number of subsystem $\mathbb{Z}_2$ line symmetries, belongs to a universal phase for MBQC \cite{RaussOkayWantStephNautrup19}.
\p The 1D cluster state was investigated in a random monitored quantum circuit context by Lavasani et al. \cite{LavasaniAlaviradBarkeshli21}. In the spirit of the cluster state belonging to a nontrivial SPT class, their circuits preserve $\mathbb{Z}_2\times\mathbb{Z}_2$ symmetry. Their three-qubit unitaries, beyond belonging to the Clifford group (which normalizes Pauli operators), furthermore conserves certain local parity operators. They found three distinct phases in their scenario, including a novel ``SPT'' phase in addition to the usual volume-law and trivial entanglement phases. Notably, they also found that all phase transitions appear to belong to the same universality class, with correlation length critical exponents of $\nu\approx 4/3$ matching $(1+1)$D percolation theory.
\p This is curious; even though the phase transition recognizes that the SPT phase differs somehow from the trivial phase, the transition in itself does not reflect the difference in computational usefulness. In a sense, the transition between volume and trivial phases is ``the same'' as the transition between the volume and SPT phases. That is to say, the transition recognizes a difference in entanglement structure apparently disjoint from a possible accompanying computational transition.
\p The question, then, arises as to whether a transition in resource power distinct from percolation can arise in random monitored quantum circuits. We might be hopeful, since the 1D cluster state is not a universal resource, whereas its 2D and 3D counterparts possess both more powerful properties and more elaborate symmetry structure. Many other resource states belong to special SPT classes with distinguished global symmetries \cite{DohertyBartlett09,Miyake10,ElseSchwarzBartlettDoherty12,MillerMiyake15,StephWangPrakWeiRaus17,RausWangPrakWeiStephen17}, but an important characteristic of the 2D cluster state is the presence of additional subsystem symmetries \cite{RaussOkayWantStephNautrup19}. Investigation of this additional level of symmetry has revealed SPT phases universal for MBQC \cite{RaussOkayWantStephNautrup19, DevakulWilliamson18,StephenNautrup19,DanielAlexMiyake20}. Enforcing such symmetries constrains random monitored circuits in a natural way, and inspires the setting of our work.

\section{$(2+1)$D Symmetric circuit}
\label{sec:2dsym}
\subsection{Setting}
\label{subsec:setting}
\begin{figure}\customlabel{fig:exa}{3(a)}\customlabel{fig:exb}{3(b)}
    \centering
    \includegraphics[width=\linewidth]{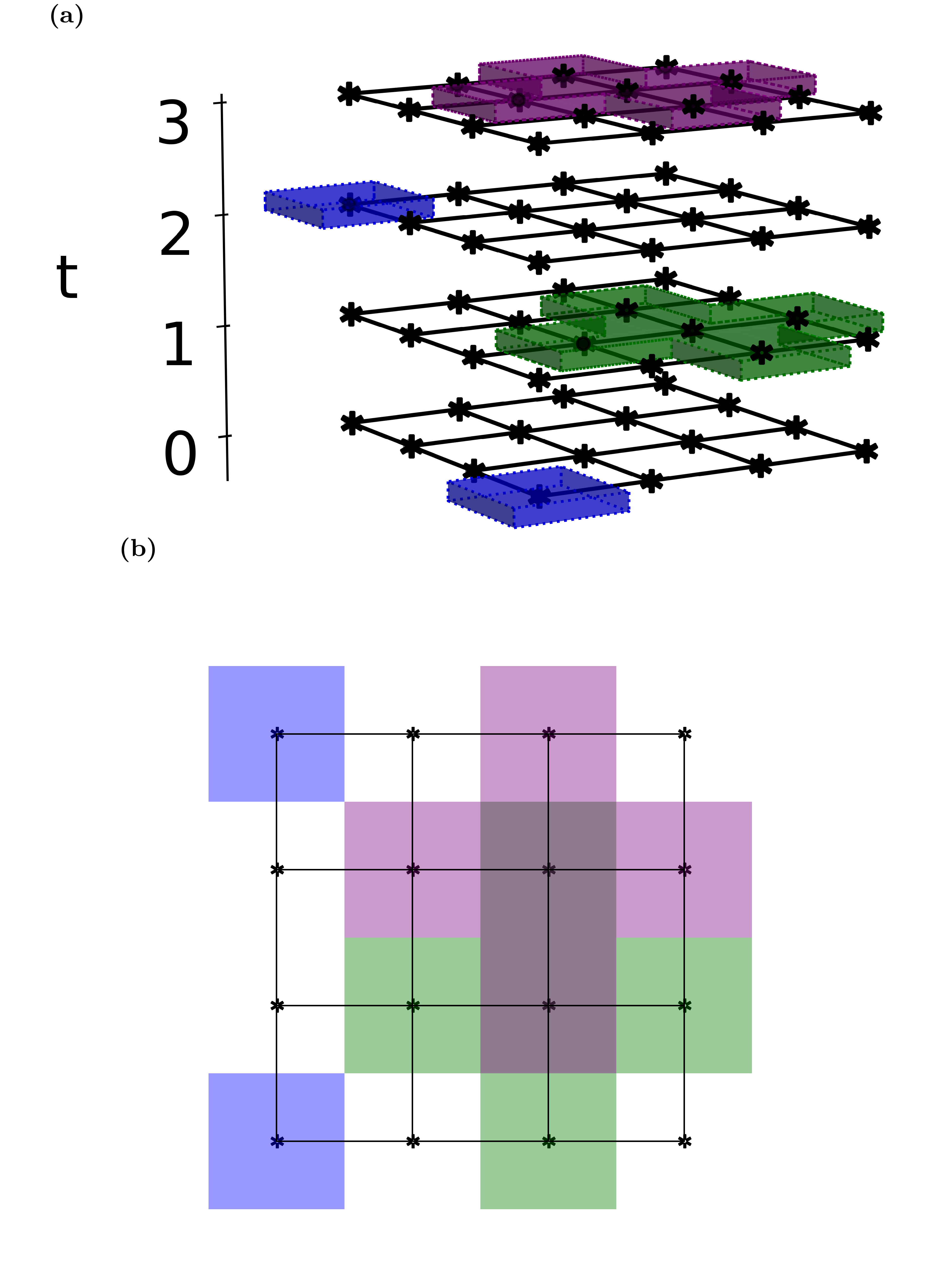}
    \caption{\label{fig:3dsample} Example 3D (a) and overhead view (b) of the circuits we consider. Two types of measurements are applied, one on a single site in the $Z$ basis and another in a ``+'' shape measuring a stabilizer of the 2D cluster state. Five-qubit unitaries are also applied in a ``+'' shape. One operation is applied every timestep $t$.}
\end{figure}

\p We study $(2+1)$D random monitored circuits with two different measurement types and draw from select unitary ensembles. Our system is defined periodically on a torus, and following the style of \cite{LavasaniAlaviradBarkeshli21} our operations applied one at a time, rather than the more standard brick-pattern (see \refig{3dsample}). At each timestep we apply one of the following operations with corresponding probabilities:

\eq{
    &U&&\text{Five-Qubit Clifford } &&p_u,\nn\\
    &M^z&&\text{Local Z Measurement} &&p^z_M,\nn\\
    &M^s&&\text{Cluster Stabilizer Measurement} &&p^s_M.
}

\begin{figure}
    \centering
    \includegraphics[width=0.8\linewidth]{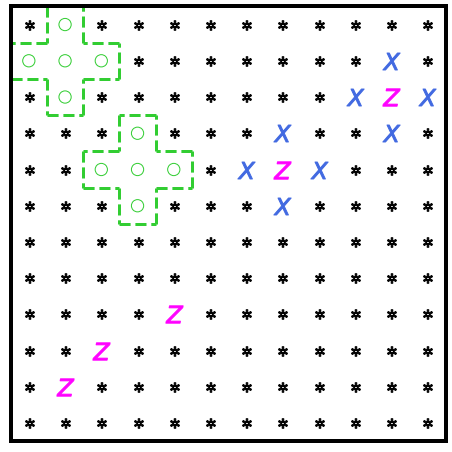}
    \caption{\label{fig:stencils} The three types of operations performed by our random circuits. Each symbol stands in for a qubit. On the lower left are local computational basis ``Z'' measurements. The upper right has measurements of 2D cluster state-like stabilizers. The upper left features five-qubit Clifford unitaries.}
\end{figure}

\p These probabilities sum to unity. Any system qubit is chosen with uniform probability to be the center of an operation. If a $Z$-basis measurement is to be performed, then the chosen qubit is measured. For the five-qubit unitary or cluster stabilizer measurement the selected qubit is the center about which the operation is applied in a ``+''-shaped stencil. This is depicted in \refig{stencils}.

\p We consider three ensembles of unitaries in our circuits. These ensembles satisfy no symmetry requirement, a global symmetry requirement, and subsystem symmetry requirements, respectively, with associated groups $\mathds{I}\subset\mathbb{Z}_2\times\mathbb{Z}_2\subset\mathbb{Z}_2^{2L-1}$ for the 2D $L \times L$ square lattice (see section 3.4 of Ref.~\cite{DanielAlexMiyake20} for an elucidation of the difference between the line symmetry $\mathbb{Z}_2^{2L-1}$ and the so-called cone symmetry $\mathbb{Z}_2^{2L}$). Recall that for more familiar settings of second-order phase transitions the Hamiltonian possesses some symmetry that is spontaneously broken across the transition, such as $\mathbb{Z}_2$ for the classical Ising model. As the Hamiltonian generates dynamics, it follows that time evolution under it observes the same symmetry. We can similarly say that a class of circuits whose constituent unitaries respect some group possesses it as a symmetry.
\p A given circuit in our setting makes use of only one ensemble of unitaries. Thus, including measurements, we have three phase diagrams corresponding to each ensemble, which with three tunable probabilities are two-dimensional. To access large system sizes via the stabilizer formalism (reviewed in Appendix~\ref{subsec:stab}) \cite{Gottesman19,AaronsonGottesman04} all ensembles are chosen to be subgroups of the five-qubit Clifford group. As our primary interest is the entanglement structure of pure states at various times, we may safely neglect stabilizer signs and the concomitant computational complexities.
\p Our first ensemble consists of the entire five-qubit Clifford group, the set of all unitaries normalizing five-qubit Pauli operators, which we say possesses the trivial symmetry group $\mathds{I}$.  The second ensemble is composed of symmetric five-qubit Cliffords respecting the global $\mathbb{Z}_2\times\mathbb{Z}_2$ checkerboard symmetry enjoyed by the 2D cluster state on a torus. This is depicted in \refig{syma}, and emulative of the symmetry considered in Ref.~\cite{LavasaniAlaviradBarkeshli21}. The symmetry's reduction to ``+''-shaped stencils can be seen in \refig{resa}. Finally, the last ensemble is constituted by Cliffords respecting the $\mathbb{Z}^{2L-1}$ subsystem symmetries of the toric 2D cluster state. These symmetries are depicted in the top subfigures of \refig{symb}, with their reduced stencils visible in \refig{resb}. Although the former two unitary ensembles are too large to store, and hence generated ``on the fly'' as needed, the last is small enough to be stored in its entirety. For more details see Appendix~\ref{subsec:generation}.

\begin{figure}\customlabel{fig:syma}{4(a)}\customlabel{fig:symb}{4(b)}
    \centering
    \includegraphics[width=\linewidth]{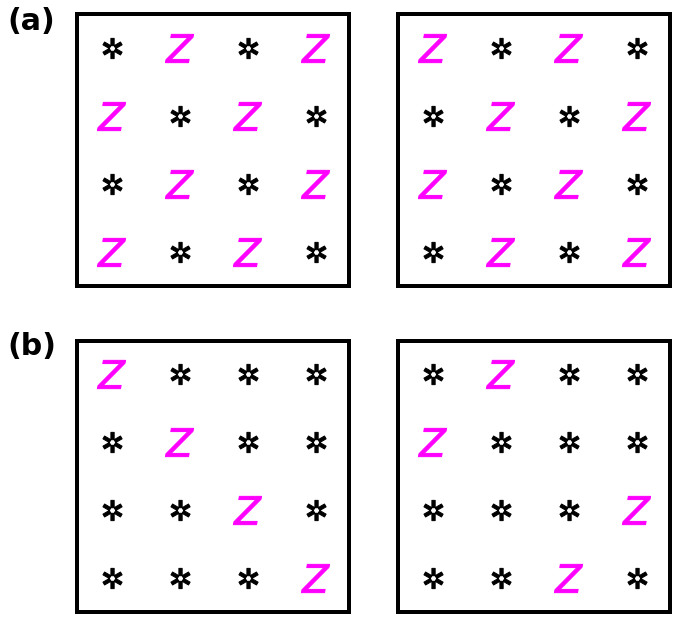}
    \caption{\label{fig:symmetries} The global $\mathbb{Z}_2\times\mathbb{Z}_2$ ``checkerboard''  (a) and subsystem $\mathbb{Z}^{\otimes 2L-1}_2$ ``diagonal'' (b) symmetries respected by Clifford ensembles. The system here consists of $4\times 4$ qubits on a periodic lattice, making the diagonal symmetries wrap.}
\end{figure}

\begin{figure}\customlabel{fig:resa}{5(a)}\customlabel{fig:resb}{5(b)}
    \centering
    \includegraphics[width=\linewidth]{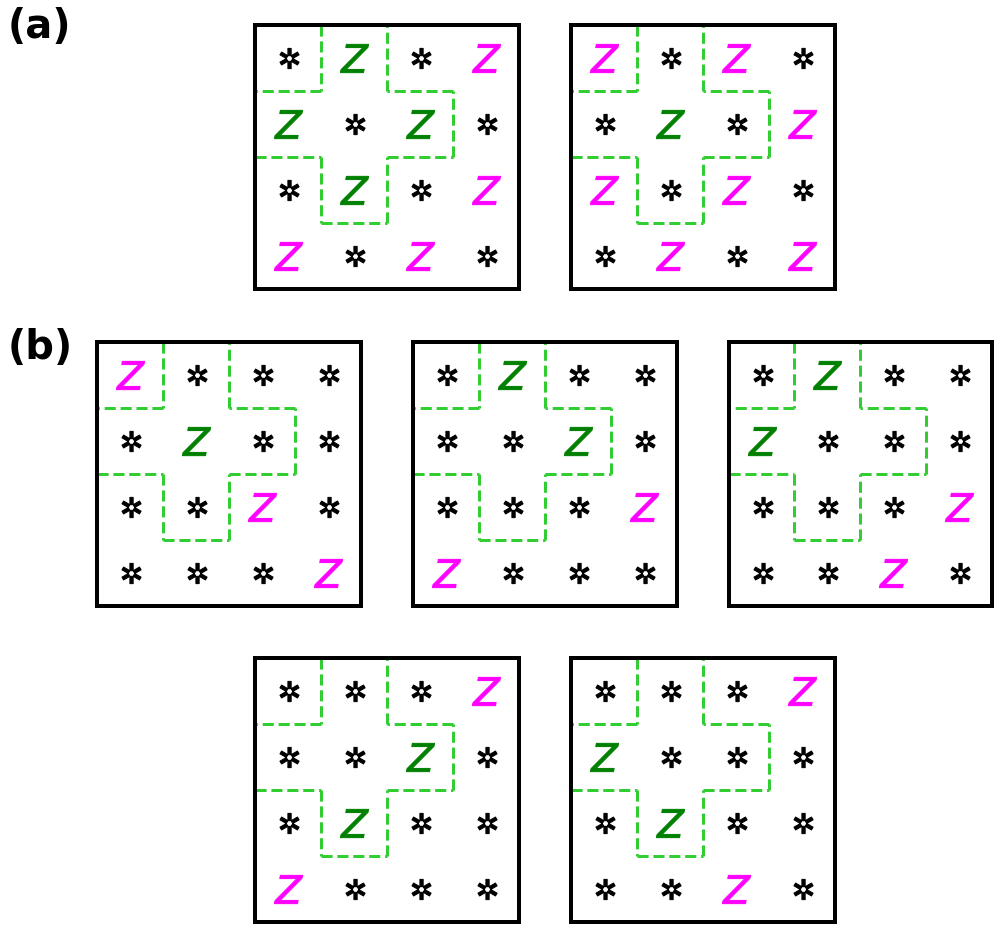}
    \caption{\label{fig:localstencils} Depiction of the Pauli operators invariant under unitaries applied within the dashed box respecting global ``checkerboard'' (a) and ``diagonal'' subsystem (b) symmetries. For the symmetry respected by an individual unitary, it suffices to reduce to the dashed boxes' contents.} 
\end{figure}

\subsection{Numerical methods}
\label{subsec:num}
\p We already stated above that we employ the stabilizer formalism and work with Clifford unitaries \cite{Gottesman19,AaronsonGottesman04}. We have written our own C99 code capable of running several simulations in parallel, managed by a Python wrapper. Our matrix manipulations are vectorized, with the binary bits of stabilizer matrices stored in 256-bit integer vectors to facilitate rapid binary operations. As phase factors are of little interest to our study, we ignore them, saving on the traditional complications of deterministic measurement-outcomes.
\p We detect transitions using three diagnostics: the topological-, ancilla-, and dumbbell entanglement entropies ($\bar{S}_{\text{top}}$, $\bar{S}_{\text{anc}}$, and $\bar{S}_{\text{dumb}}$) \cite{KitaevPreskill06,LevinWen06,GullansHuse20p,GullansHuse20,WilliamsonDuaCheng19}. The bar formally denotes averaging over both circuit realizations and measurement outcomes. $\bar{S}_{top}$ is our primary workhorse. We make use of the seven-term variant in Eq.~(\ref{eq:top}) with regions defined as in \refig{q1dregions} along quasi-1D cylinders. We have furthermore cross-checked our analysis with other geometries, such as diagonal subregions, as well as a more traditional ``ring'' setup. Appendix~\ref{subsec:topologicalentanglement}, besides discussing the finer details of our implementation, also shows that the critical points and exponents match within numerical error for these other geometries. The ancilla entanglement entropy $\bar{S}_{\text{anc}}$ has its value as a geometrically independent sanity check, and our implementation is discussed in Appendix~\ref{subsec:ancilla}.
\p This question of geometry is relevant, due to the effect of so-called spurious topological entanglement entropy \cite{WilliamsonDuaCheng19,StephenDreyerIqbalSchuch19}, wherein subsystem symmetries may induce additional correlations, which the topological entanglement entropy defined for special regions may be sensitive to. The dumbbell entanglement entropy \cite{WilliamsonDuaCheng19}, which is a diagnostic of subsystem symmetry, turns this inconvenience into a tool. We employ it to detect the transition in the case of pure measurement dynamics (i.e., when $p_u=0$), where neither the topological entanglement entropy nor the ancilla entanglement entropy are effective. The details are contained in Appendix~\ref{subsec:dumbbell}.
\p For finite-size scaling analysis we employ three different curve-fitting schemes to retrieve details of our transitions . The first method, ``Nearest,'' seeks to make a hypothesized common curve connecting adjacent scaled points along the common curve as smooth as possible. The second, ``Multilevel,'' compares not with just the nearest scaled points, but all the nearest scaled points for all different system sizes. Finally, the last, ``Polynomial,'' is as the name suggests a high-order polynomial fit quantified by the residual. For sufficiently well-behaved data and enough datapoints the three typically agree, although the second would seem the most robust, while the last produces the smoothest plots. More in-depth coverage of this topic may be found in Appendix~\ref{sec:fit}.

\begin{figure*}[!thp]\customlabel{fig:regphase}{6(a)}\customlabel{fig:sptphase}{6(b)}
    \centering
    \includegraphics[width=\linewidth]{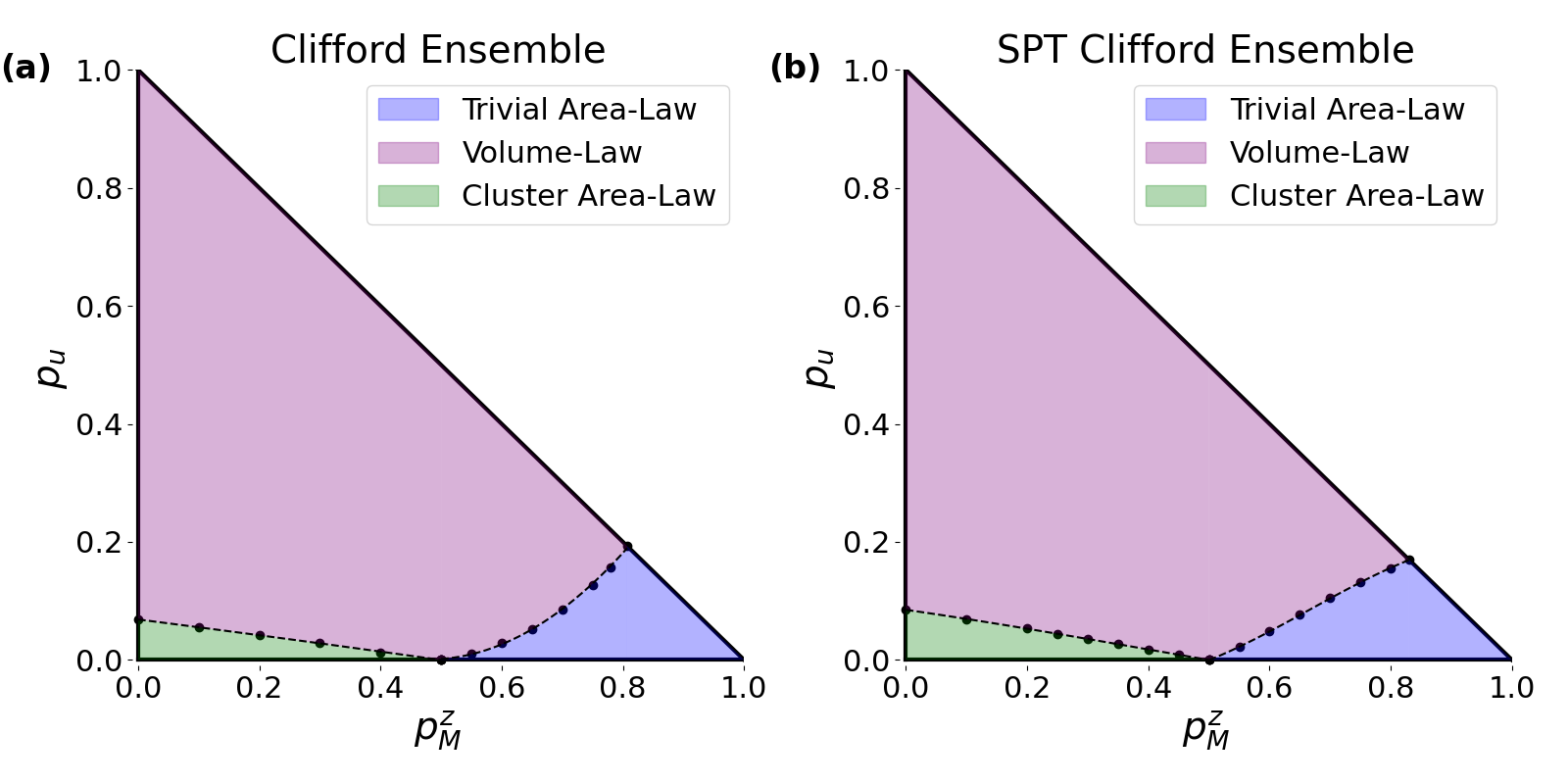}
    \caption{\label{fig:boringphases} Phase diagrams for the unconstrained Clifford ensemble (a) and global symmetry-constrained (b) ensembles in terms of the probabilities of applying five-qubit Cliffords $p_u$ and computational basis $Z$ measurements $p_M^z$. Phase transitions occur at black dots, interpolations between which form the phase boundaries. Both diagrams are qualitatively similar, with small area-law entanglement regions due to the scrambling power of the five-qubit Clifford ensembles.}
\end{figure*}

\subsection{Results}
\label{subsec:results}
\subsubsection{Unconstrained Clifford group}
\label{subsubsec:reg}
\begin{figure*}[!htbp]\customlabel{fig:7a}{7(a)}\customlabel{fig:7b}{7(b)}
\centering
    \includegraphics[width=0.95\linewidth]{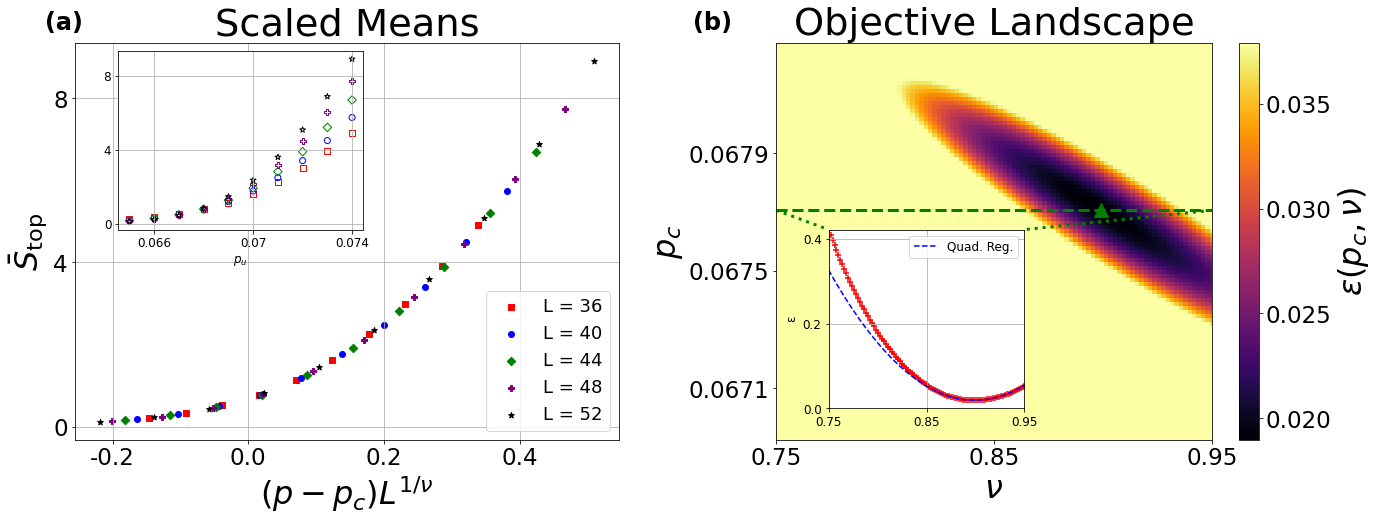}
    \caption{\label{fig:sptvolregq1d} Data/analysis using the circuit/trajectory-averaged seven-term topological entanglement entropy $\bar{S}_{\text{top}}$ with quasi-1D cylindrical geometry for the cluster-to-volume transition for the unconstrained Clifford ensemble. (a) Collapsed data as a function of the probability of applying a unitary operation $p_u$ (unscaled data inset). (b) Objective landscape for the collapse using a polynomial fitting method (slice at $p_u=p_c$ inset). The minimum is found at the critical point $p_c\approx 0.068$ and critical exponent $\nu\approx 0.90$.}
\end{figure*}

\begin{figure*}[!htbp]\customlabel{fig:8a}{8(a)}\customlabel{fig:8b}{8(b)}
\centering
    \includegraphics[width=0.95\linewidth]{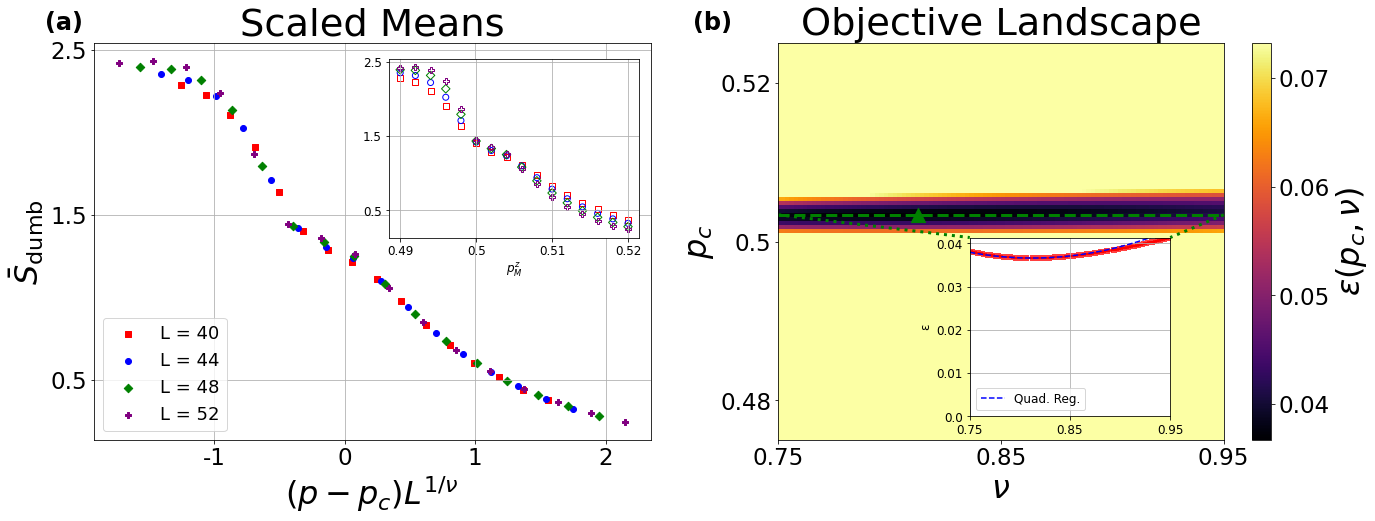}
    \caption{\label{fig:puremeasurement} Data/analysis using the circuit/trajectory-averaged dumbbell entanglement entropy $\bar{S}_{\text{dumb}}$ for pure measurement dynamics ($p_u=0$). No unitaries are employed, so this is the same for every ensemble. (a) Collapsed data as a function of the probability of performing a local computational basis measurement with tuning parameter $p=p_M^z$ (unscaled data inset). (b) Objective landscape determining the collapse (slice at $p_M^z=p_c$ inset). The minimum is at critical point/exponent $p_c\approx 0.5$/$\nu\approx 0.84$.}
\end{figure*}

\p Our unconstrained Clifford group refers to the group of all five-qubit Cliffords. Its symmetry group may be said to be trivial. The entire phase landscape is shown in \refig{regphase}. There are three phases: a trivial area-law phase, a cluster area-law phase, and one volume-law phase. As a consequence of the generic entangling power of five-qubit Cliffords, the volume region is considerably larger than the other two.
\p The critical behavior along the $p_M^s=0$ line (where no cluster stabilizers are measured) is expected \cite{VasseurFisherLudwig21}, and has been numerically shown \cite{SierantSchiroLewensteinTurkeshi22} (with some prior controversy \cite{TurkeshiFazioDalmonte20,LuntSzynisweskiPal21}), to exhibit behavior akin to the 3D bond percolation transition. In particular, the correlation critical exponent is expected to take on the value $\nu\approx 0.88$. It is maybe not surprising that this critical behavior persists throughout the phase diagram---that is, even when we introduce cluster stabilizer measurements. 
\p We observe such a similar critical exponent along the $p_M^z=0$ line. The data, along with the fitting landscape and the finite size scaling associated with the minimum thereof, is depicted in \refig{sptvolregq1d}, where $\nu\approx 0.90$. This is not surprising for the reason that the essential phenomena seen in the one-operation-per-time-step scheme we employ do not differ from more standard maximally scrambling brickwork setup. In these latter setups, however, the unitary subgroup averaging property may be used to replace single-qubit measurements by another operator sharing the same locality as the applied unitaries. Specifically, since the Clifford unitary taking (in the Heisenberg picture) the operator $Z$ to the cluster state stabilizer $XXZXX$ belongs to the set of all five-qubit Cliffords, it follows that we may heuristically exchange the roles of measuring either operator and expect qualitatively similar phenomena. It is important to note, however, that this argument cannot be applied for our SPT or SSPT ensembles, as symmetry demands that all permitted unitaries take the central $Z$ operator to itself.
\p We also include in this subsubsection analysis of pure measurement dynamics, which, being in the $p_u=0$ limit where no unitaries are applied, is the same for all ensembles. The data is depicted in \refig{puremeasurement}. Numerical collapse shows the critical point to be at about $p_c\approx 0.5$, which a duality argument similar to Refs.~\cite{LavasaniAlaviradBarkeshli21,LavasaniAlaviradBarkeshli21p} suggests is exact. The critical exponent, meanwhile, is $\nu\approx 0.85$, which again matches $(2+1)$D percolation. Further investigation into the properties of this pure measurement critical point are found in Appendix~\ref{sec:critmeas}. Between all of these, we have ample evidence that the criticality class is constant along all phase boundaries for this unconstrained ensemble of five-qubit Clifford unitaries. 

\subsubsection{SPT Clifford group}
\label{subsubsec:spt}
\p The SPT Clifford group considered consists of all five-qubit Cliffords respecting the $\mathbb{Z}_2\times\mathbb{Z}_2$ checkerboard symmetry group of the 2D cluster state. This is akin to a restriction placed on three-qubit Cliffords on symmetric $(1+1)$D circuits, which has a phase characterized by measurements of 1D cluster state stabilizers \cite{LavasaniAlaviradBarkeshli21}. The global symmetries in question preserved by five-qubit unitaries are depicted in \refig{syma}, with associated stencils in \refig{resa}.
\p Our results in this case are similar to Ref.~\cite{LavasaniAlaviradBarkeshli21}. We find that although three distinct phases exist, the critical exponent remains close to that of $(2+1)$D percolation, with $\nu\approx 0.84$. The phase diagram, presented in \refig{sptphase}, is strikingly similar to the unrestricted ensemble above. The primary visual difference is slight growth in the size of the trivial area-law phase, owing to the imposed symmetry restrictions making the unitary ensemble somewhat less entangling.

\subsubsection{SSPT Clifford group}
\label{subsubsec:sspt}

\begin{figure}[!hbp]
    \centering
    \includegraphics[width=0.9\linewidth]{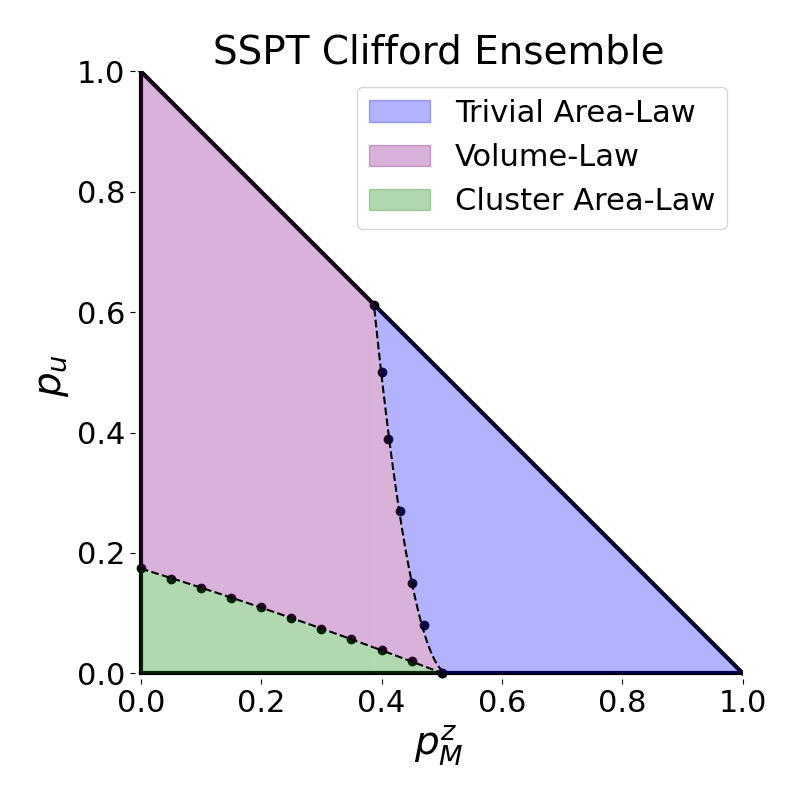}
    \caption{\label{fig:ssptphase} Phase diagram for the SSPT Clifford ensemble in terms of the probability of applying a 5-qubit unitary $p_u$ and a computational basis $Z$ measurement $p_M^z$. The three phases are separated by boundary lines determined by interpolating critical points obtained numerically at the black dots. Compared to \refig{regphase} and \refig{sptphase}, the trivial and cluster area-law regions are considerably larger, and the critical exponents generally differ.}
\end{figure}

\begin{figure*}[!htp]\customlabel{fig:10a}{10(a)}\customlabel{fig:10b}{10(b)}
\centering
    \includegraphics[width=0.95\linewidth]{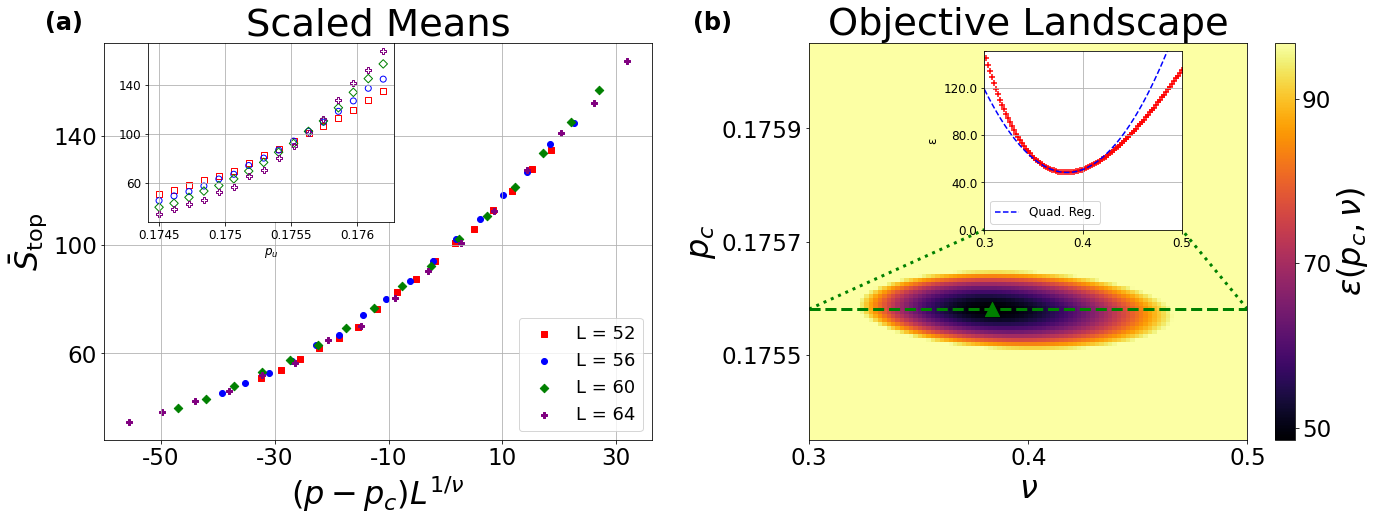}
    \caption{\label{fig:sptvolmaxq1d} Data/analysis using the circuit/trajectory-averaged seven-term topological entanglement entropy $\bar{S}_{\text{top}}$ with quasi-1D cylindrical geometry for the cluster-to-volume transition for the SSPT Clifford ensemble. (a) Collapsed data as a function of the probability of applying a unitary operation $p_u$ (unscaled data inset). (b) Objective landscape for the collapse using a polynomial fitting method (slice at $p_u=p_c$ inset). The minimum is found at the critical point $p_c\approx 0.176$ and critical exponent $\nu\approx 0.38$.}
\end{figure*}

\begin{figure*}[!htp]\customlabel{fig:11a}{11(a)}\customlabel{fig:11b}{11(b)}
    \centering
    \includegraphics[width=0.95\linewidth]{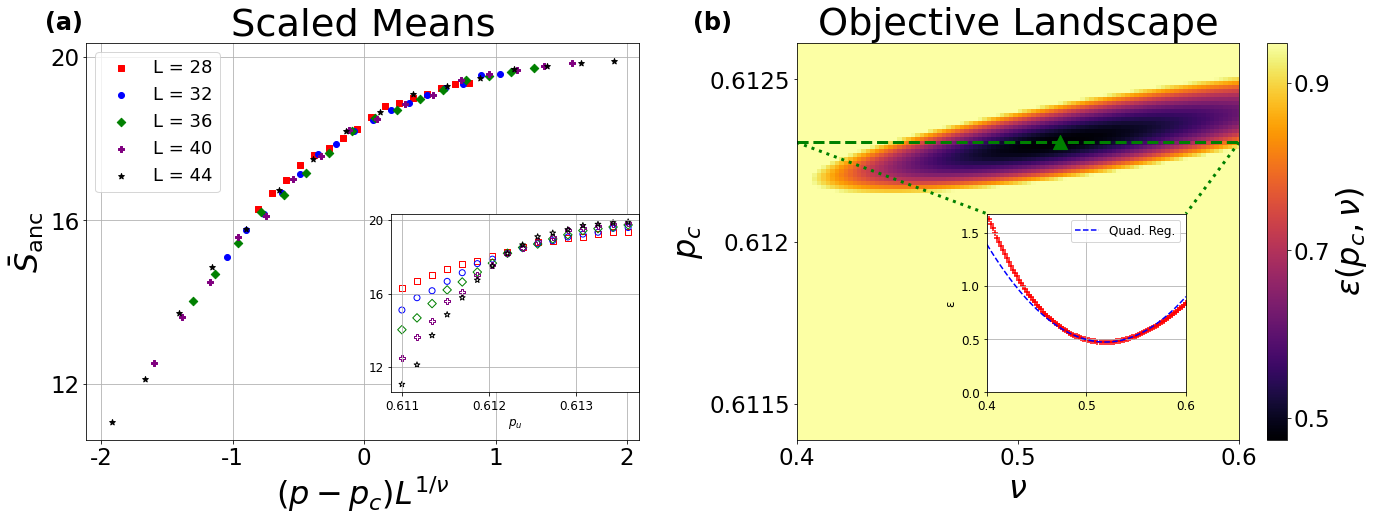}
    \caption{\label{fig:trivvolanc} Data/analysis using the circuit/trajectory-averaged ancilla entanglement entropy $\bar{S}_{\text{anc}}$ for the volume-to-trivial-area-law transition. (a) Collapsed data as a function of the probability of applying a unitary operation $p_u$ (unscaled data inset). (b) Objective landscape determining collapse (slice at $p_u=p_c$ inset). The minimum is at critical point/exponent $p_c\approx 0.612$/$\nu\approx 0.52$.}
\end{figure*}

\p The SSPT Clifford group is constituted by the five-qubit Cliffords preserving the $\mathbb{Z}^{2L-1}_2$ line symmetries of the 2D cluster state. This includes, as a subgroup, the symmetries respected by the SPT ensemble, and so this constraint is strictly stronger. Examples of the line symmetries are depicted in \refig{symb}, while the local stencils are shown in \refig{resb}. For the 2D cluster state on open boundaries, as with global symmetry of the 1D cluster state, these symmetries act projectively on edge modes, protecting boundary degeneracy \cite{YouDevakulBurnellSondhi18}. Although our dynamics take place on the torus, we nevertheless preserve these symmetries, and hence expect to preserve at least some of the resource's special properties.
\p The phase landscape is depicted in \refig{ssptphase}. The qualitative features of the three regions have noticeably changed compared to the previous two ensembles. The area-law regions have expanded considerably, owing to our stringent constraints severely hampering the entangling capabilities of viable unitaries. The trivial entanglement phase in particular has grown, as not only do all unitaries preserve the Pauli $Z$ acting on the central qubit, but the individual subsystem symmetries further fix the output of every $Z$ to only a few dozen possibilities.
\p We find that the critical exponent along the $p^z_M=0$ line differs significantly from the percolation exponent. Through several diagnostics, and three separate fitting routines, we consistently find $\nu\approx 0.39$. Example data is depicted in \refig{sptvolmaxq1d}---similar results for different $\bar{S}_{\text{top}}$ geometries are found in Appendix~\ref{subsec:topologicalentanglement}. Comparable values of $\nu$ are found along the phase boundary, terminating at the $p_u=0, p^z_M=0.5$ measurement-only critical point. The volume-to-trivial-area-law transition for this ensemble is presented in terms of the ancilla entanglement in \refig{trivvolanc}, with $\nu\approx 0.52$. Both critical exponents differ considerably from the $(2+1)$D percolation universality class with its characteristic $\nu\approx 0.87$.
\p We interpret these new exponents to reflect a shortfall of the previously known percolation mapping \cite{SkinnerRuhmanNahum19,Skinner23}, wherein all unitaries becomes impasses regardless of their original form. On the one hand, the mapping is certainly justifiable for Haar-random unitaries---intuitively, the unitaries that could be cut through, like the set of products of single-qubit unitaries, constitutes a set of measure zero. For Clifford-random, and even for 1D $\mathbb{Z}_2\times\mathbb{Z}_2$-respecting Cliffords \cite{LavasaniAlaviradBarkeshli21}, this mapping still appears empirically justified, possibly because typical unitaries in these ensembles remain sufficiently entangling. On the other hand, for our final ensemble here, the large number of restrictions for the subsystem symmetries we have imposed---especially the preservation of the center qubit's Pauli $Z$---may make this simplification no longer applicable.

\subsubsection{Volume Phase Suppression}
\label{subsubsec:suppress}
\begin{figure}[hb]\customlabel{fig:suppsyma}{12(a)}\customlabel{fig:suppsymb}{12(b)}
      \centering
      \includegraphics[width=\linewidth]{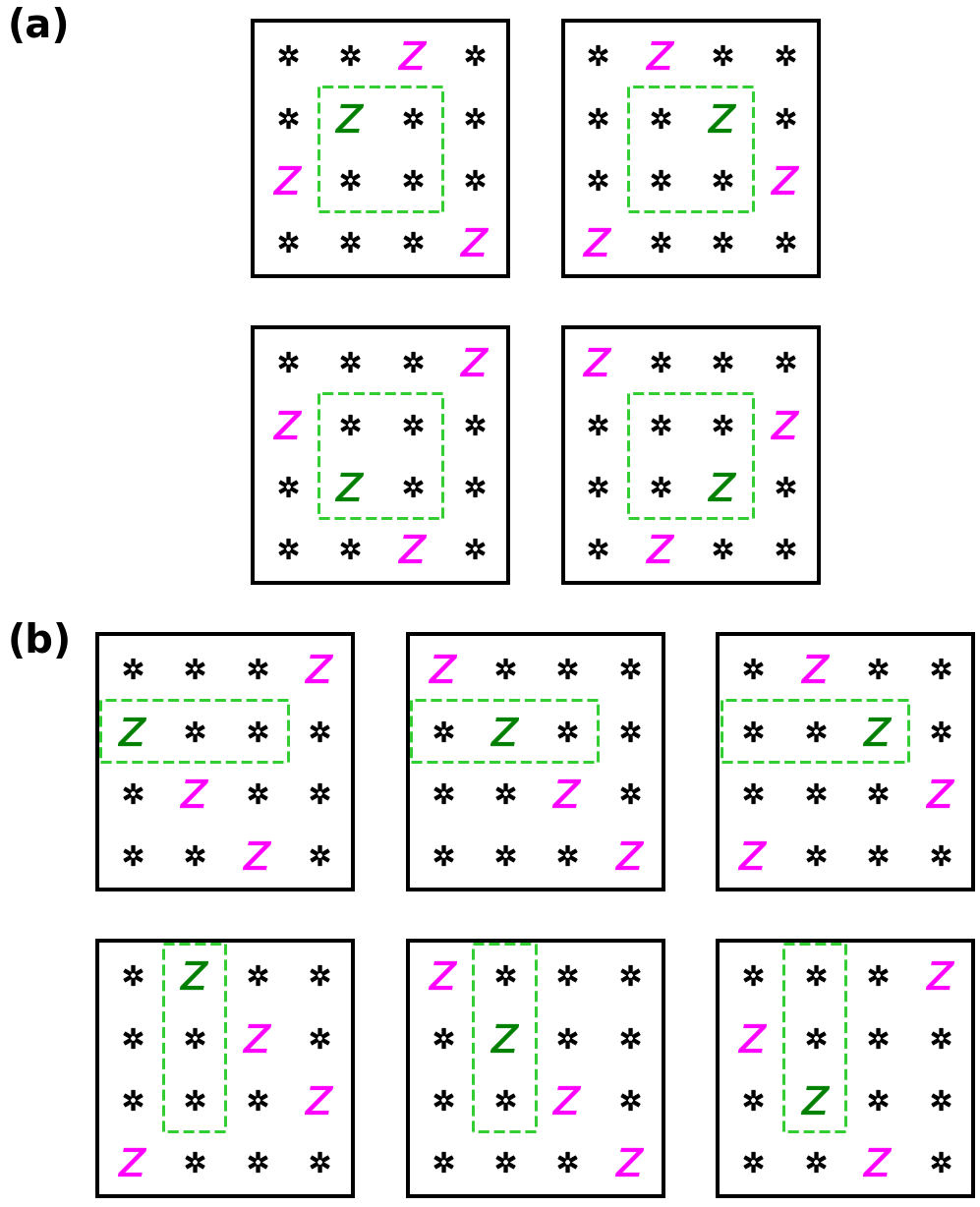}
      \caption{\label{fig:suppsymm} Illustration of the Pauli operators preserved by a unitary applied within the dashed box for ``diagonal'' subsystem symmetries for four (a) and three (b) qubits. As the unitary acts trivially outside the box, the subsystem symmetry of \refig{symb} requires local conservation of each Pauli $Z$.} 
\end{figure}
\begin{figure*}[ht]\customlabel{fig:refphasea}{13a}\customlabel{fig:refphaseb}{13b}
      \centering
      \includegraphics[width=0.9\linewidth]{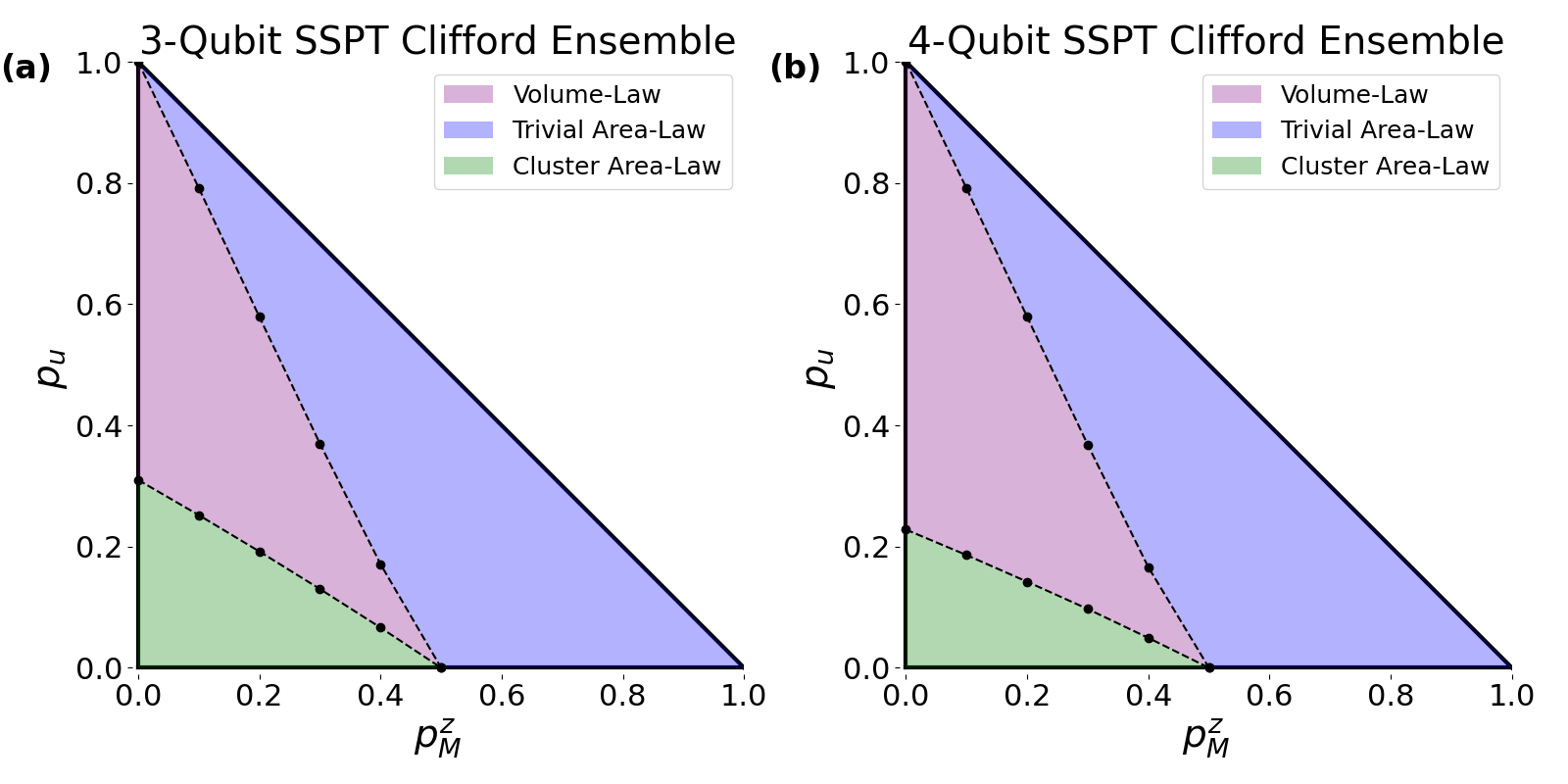}
      \caption{\label{fig:refphase} Phase diagrams for three-qubit (a) and four-qubit (b) subsystem-symmetric unitary ensembles. Unitaries, computational basis $Z$ measurements, and stabilizer measurements are applied with probabilities $p_u, p_M^z,$ and $p_M^s$ subject to unit sum. The phase boundaries are determined by interpolation using numerical simulations at the black dots. Local charge conservation causes the volume-law phase to be suppressed even more than in \refig{ssptphase}.}
\end{figure*}

\p A remarkable feature of our subsystem-symmetric ensemble of circuits is its comparatively reduced volume-law phase. This suppression of volume-law behavior is of some interest, given early arguments that it should vanish as the system size grows \cite{ChanNandkishorePretkoSmith19}, and its notable absence in select systems \cite{CaoTilloryDeLuca19}. We may make the suppression even more dramatic in our case by considering subsystem-symmetric Cliffords operating on fewer qubits. For four qubits in a square or three on a line (see \refig{suppsymm}) subsystem symmetry necessitates that every unitary preserve the Pauli $Z$ of each individual qubit. The associated phase diagrams are shown in \refig{refphase}. Therein, not only is the cluster area-law phase quantitatively larger compared to \refig{ssptphase}, but the transition has completely disappeared from the computational basis measurement/unitary line $p_M^s=0$. Indeed, the trivial area-law phase now takes up half the phase diagrams in \refig{refphase}.  
\p We can understand the growth of area law phases as follows. The full $n$-qubit Clifford group can be generated by the phase $S_i$ and Hadamard $H_i$ gates for every qubit, and the controlled-$Z$ $CZ_{ij}$ between every two qubits. If we demand that our unitaries commutate with every Pauli $Z_i$, we obtain the Abelian subgroup generated by all $S_i$'s and $CZ_{ij}$'s (this group corresponds to $2^{\tfrac{n(n+1)}{2}}$ symplectic matrices for our sign-agnostic purposes). This unitary subgroup has rather limited entangling capabilities, and furthermore the symmetry has the important consequence that all elements also commutate with $Z$ measurements. Hence, for a given circuit, subsystem-symmetric unitaries may be commutated cleanly past any $Z$ measurements, and through 2D cluster stabilizer measurements at the cost of mild operator growth. Once all unitaries are moved to the start of the circuit---which then operate either trivially on a product initial state or may be absorbed into a Clifford random input---the subsystem-symmetric circuit can be reframed as a competition between local $Z$ measurements and metastasized cluster state stabilizers.
\p The fact that the trivial area-law to volume-law phase transition now occurs very close to the $p_M^z=p_M^s$ line is natural from this perspective. The trivial area-law phase can be extrapolated from the case of pure competing measurements where $p_u=0$ and $p_M^z>0.5$ to $p_M^z\gtrapprox p_M^s$ more generally. We can rephrase this observation that since all symmetric unitaries possess the trivial all-$Z$ stabilizer product state as an eigenvector, it is the rate of cluster state measurements $p_M^s$ that determines what phase the circuit belongs to. While there is no such common eigenvector for the entirety of our five-qubit SSPT ensemble, the high symmetry restrictions have a preference for low-entanglement states. This would seem to be a key element underlying the suppression of volume-law behavior.

\section{Discussion}
\label{sec:discussion}
\begin{table*}
    \begin{center}
        \begin{tabular}{ l l  r  r r r  r r }
            \hhline{========}
              &\multirow{2}{*}{\backslashbox{Symmetry}{Types}} & &\multicolumn{2}{ c }{Volume/Cluster Area } & &\multicolumn{2}{ c }{Volume/Trivial Area}   \\
            Ensemble& & & \multicolumn{2}{c}{($p_c, \nu$)} & &\multicolumn{2}{c}{($p_c, \nu$)} \\ \hline
            Clifford& $\mathds{I}$ & &$0.068\pm0.001$ & $0.90\pm0.03$ & &$0.193\pm 0.005$& $0.85\pm0.03$     \\
            SPT Clifford& $\mathbb{Z}_2\times\mathbb{Z}_2$ & &$0.084\pm 0.002$  & $0.83\pm 0.03$ &  &$0.170\pm0.001$   & $0.83\pm0.05 $       \\
            SSPT Clifford& $\mathbb{Z}_2^{2L-1}$ & &$0.176\pm 0.001$ & $0.38\pm 0.05$   & &$0.612\pm 0.002$ & $0.52\pm0.06 $             \\
            \hhline{========}
        \end{tabular}
        \caption{\label{tab:exp} Critical points $p_c$ and exponents $\nu$ along two phase boundaries. $p_c$ refers to the value of $p_u$ with $p_M^z=0$ for the volume/cluster area-law and $p_M^z=1.0-p_u$ for the volume/trivial area-law phase transitions, respectively. Error bars are determined by fitting an objective function about a minimum and taking the width of twice the minimum.}
    \end{center}
\end{table*}
\p We have studied the interplay between a hierarchy of increasingly restrictive symmetries and the phases observed in $(2+1)$D random monitored quantum circuits. We have shown that, in contrast to the globally symmetric ensembles in (1+1)D \cite{LavasaniAlaviradBarkeshli21}, the transitions of random monitored quantum circuits with a subsystem-symmetric ensemble of Clifford unitaries features a different universality class than unrestricted dynamics in $(2+1)$D. Our results are summarized in Table~\ref{tab:exp}. The volume-to-cluster-area-law transition has a critical exponent of $\nu\approx 0.38$, while the volume-to-trivial-area-law transition has $\nu \approx 0.52$. These values are quite different from $\nu\approx 0.88$ for globally symmetric and unrestricted five-qubit Cliffords. The (1+1)D SPT case is analogous to the global symmetry considered here, where the transitions all appear to remain within the percolation universality class. This suggests that the imposition of subsystem symmetry fundamentally alters certain critical dynamics.
\p By considering diagnostics besides $\bar{S}_{\text{top}}$ we can understand how this difference in $\nu$ and the criticality class is not merely cosmetic, but has a certain operational significance in terms of the system's response to single-qubit measurement. Variants of the ancilla entanglement entropy \cite{GullansHuse20,GullansHuse20p} (see Appendix.~\ref{subsec:ancilla}) give us insight into how quantum information can be propagated across time (and by duality at criticality, space \cite{Google23a,BaoBlockAltman24}) in monitored systems. In the volume-law phase, where information is scrambled, entangled ancilla qubits retain correlations with the system for a long time. In contrast, quantum information is quickly read out by the monitored circuit in the trivial area-law phase. The rate at which the survival time increases or decreases in each respective phase as system size increases is determined by $\nu$.
\p More concretely, the scaling hypothesis dictates that the relevant timescale goes as $\tau\sim|p-p_c|^{-\nu}$ \cite{GullansHuse20p}. By duality at criticality, we expect the units to be given by some monotonic function $f$ of $L$, so the scaling is $\tau(L,p)\sim|p-p_c|^{-\nu}f(L)$. This timescale may be, for example, the survival time $\tau$ of exponential decay $e^{-t/\tau}$, but as the only relevant scale all temporal quantities should behave similarly. By keeping $p-p_c$ constant, and increasing $L$, we see that this relevant scale grows more rapidly for larger $\nu$. Operationally this translates to survival times being larger in the volume-law phase for larger $\nu$. In this sense, near criticality, the characteristic phenomena of the distinct phases are more pronounced for larger $\nu$. This matches the intuition that the unconstrained Clifford ensemble has greater scrambling power, and hence the properties of the two phases are more pronounced.

\begin{figure*}\customlabel{fig:telanca}{14(a)}\customlabel{fig:telancb}{14(b)}\customlabel{fig:telancc}{14(c)}\customlabel{fig:telancd}{14(d)}
    \centering
    \includegraphics[width=0.85\linewidth]{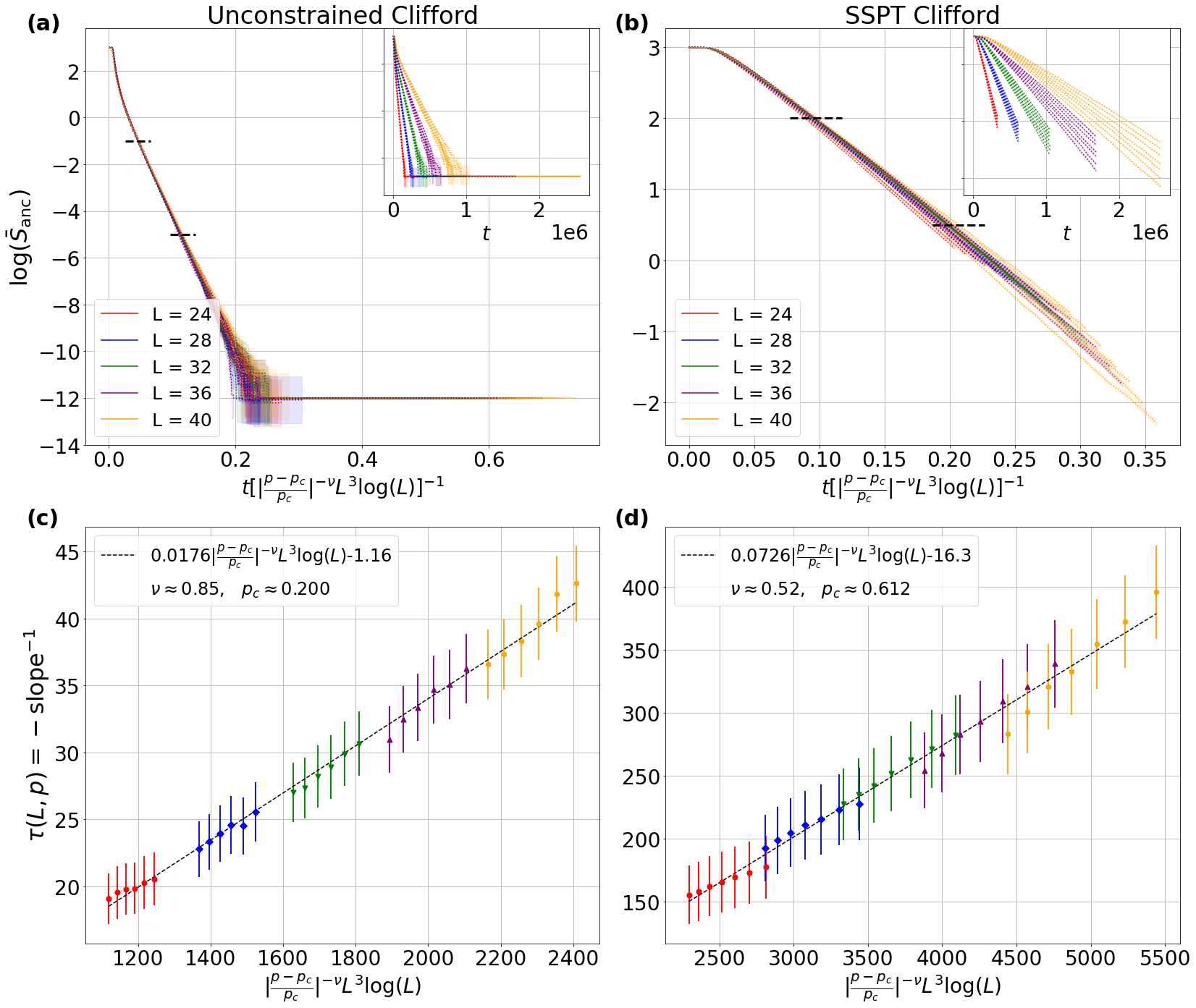}
    \caption{\label{fig:telanc} Upper: Logarithm of the ancilla entanglement entropy slightly beneath criticality in the area-law phase for unconstrained (a) and SSPT (b) Cliffords for different system sizes $L$ and tuning parameters $p=p_u=1-p_M^z$. All measurements are single-qubit ($p_M^s=0$). The main panels show how the inset curves overlap when expressed as a function of the normalized time $t(|p-p_c|^{-\nu}L^3\log(L))^{-1}$. Lower: behavior of the survival time $\tau$, the negative reciprocal of the slope of $\log(\bar{S}_{\text{anc}})$, for the same subcritical unconstrained (\refig{telancc}) and SSPT (\refig{telancd}) Clifford circuits. The slopes are taken between the dashed black lines of the top panels. The hypothesis that the relevant time scale goes as $|p-p_c|^{-\nu}L^3\log(L)$ yields a respectable regression. We impose a floor to the logarithm function in the unconstrained case. }
\end{figure*}

\p We can be quantitative by looking at the logarithm of $\bar{S}_{\text{anc}}$ over time for the unconstrained and SSPT Clifford ensembles. We consider, with particular mind towards the single-qubit measurements of MBQC, subcritical trivial area-law circuits near the transition to the volume-law phase. There are no cluster state measurements $(p_M^s=0)$, so our tuning parameter obeys $p=p_u=1-p_M^z$--- this corresponds to the diagonals of \refig{regphase} and \refig{ssptphase}. 
\p The insets of the upper panels of \refig{telanc} show, for a variety of different system sizes $L$ and tuning parameters $p$, linear behavior in $\log(\bar{S}_{\text{anc}})$ after a time. That is, they show how the decay is asymptotically exponential: $\bar{S}_{\text{anc}}\propto e^{-t/\tau(L,p)}$. Notably, the unconstrained Clifford ensemble features quadratic behavior eventually giving way to linearity, whereas the subsystem-symmetric ensemble is essentially linear after the initial plateau. The upper panels of \refig{telanc} show how these lines collapse to a common curve as a function of $t[|p-p_c|^{-\nu}L^3\log(L)]^{-1}$. In fact, $\nu$ and $p_c$ may be inferred by optimizing with respect to this collapse, although this method is not to be recommended. Using this quasi-exponential behavior $\bar{S}_{\text{anc}}\propto e^{-t/\tau(L,p)}$, the lower figures further show how the survival times $\tau$---the negative reciprocals of these lines' slopes---behave as a function of $p-p_c$ and $L$. Again, we can optimize with respect to $p_c$ and $\nu$ to recover critical quantities, and again we empirically find that a useful timescale is given by $|p-p_c|^{-\nu}L^3\log(L)$. This curious form of $L$ scaling arises from a coupon collecting argument (see, e.g., Ref.~\cite{ErdosRenyi61}). Our simulations implement one operation per timestep, so what we call $t$ is on the order of $\tilde{t}L^2\log(L)$, where $\tilde{t}$ is the number of timesteps in a parallelized circuit; the scaling $\tilde{t}|p-p_c|^{\nu}/L$, then, is a more natural-looking hypothesis, especially since parallelized circuits map better to statistical models \cite{SkinnerRuhmanNahum19,ChoiBaoQiAltman20,VasseurFisherLudwig21}. The fact that $\nu$ may be hypothetically obtained this way, which reflects a difference in near-critical decay behavior, shows how distinct critical behaviors have a functional impact on temporal correlations.
\p This analysis also has some bearing on our MBQC inspiration. Near criticality, we expect a duality of spatial and temporal behaviors by virtue of a conformal field theory. The behavior of $\bar{S}_{\text{anc}}$, which quantifies entanglement over time, should be similar to measures of entanglement carried over space, which evokes the long-range entanglement that is a necessary (though not sufficient) prerequisite for MBQC. This translates to the intuitive statement that the unconstrained Clifford ensemble approaches volume-law behavior at a quicker rate compared to the SSPT Clifford ensemble and also builds up more long-range entanglement. The relevant fact here is that a surplus of entanglement is useless for MBQC (though we should distinguish that our volume-law phase is Clifford- rather than Haar-random) \cite{GrossFlammiaEisert09,BremnerMoraWinter09}. That is to say, not only does the mere presence of subsystem symmetry enforce a sufficient amount of the ``right'' kind of entanglement for MBQC, but additionally the fact that the associated unitary ensemble is less entangling also inhibits the approach to excessive volume-law behavior.
\p It is also an interesting pursuit to compare our setting with other hierarchies of symmetries in phase transitions. Recall that our (2+1)D Clifford, SPT Clifford, and SSPT Clifford circuits feature symmetry groups $\mathds{I}\subset \mathbb{Z}_2\times\mathbb{Z}_2\subset \mathbb{Z}_2^{2L-1}$, respectively. The $O(N)$ classical spin model, whose Hamiltonian is $H=J\sum_{\langle \alpha,\beta\rangle}\mbf{S}_\alpha\cdot\mbf{S}_\beta$ for the $N$-component spin, contains the Ising, XY, and Heisenberg models as special cases, with associated symmetry groups $\mathbb{Z}_2\subset O(2)\subset O(3)$, respectively. The critical phenomena of these models have been studied extensively, and the critical exponent $\nu$ is known to increase with $N$ in this hierarchy---approximately equal to $0.63, 0.67,$ and $0.70$, respectively in the spatial dimension $d_s=3$ \cite{GorLarTka84,GuidaZinn98,Hasenbusch01}. It is believed that this trend is true regardless of whether the spin is treated classically or quantumly. It is also known that the one-loop analysis of the fixed point in more general spatial dimension $d_s=4-\epsilon$ theory yields the following expression for the exponent for general $N$:
\eq{
    \frac{1}{\nu}=2-\frac{N+2}{N+8}\epsilon.\nonumber
}
This suggests that $\nu$ generally increases with $N$ \cite{CardyScale}. 
\p In contrast, for our hierarchy of lattice symmetries, $\nu$ is about the same for $\mathds{I}$ and $\mathbb{Z}_2\times\mathbb{Z}_2$ ($0.83$ and $0.85$), and decreases for the subsystem symmetry to $0.38$. This is the opposite trend. It might be fruitful to investigate the relationship between hierarchies of symmetries and critical exponents in broader contexts.
\p One final matter to note is that, just as for the full Clifford group, the critical exponents observed for the SSPT ensemble violate the Harris criterion---in fact, the violation is even more severe \cite{Harris74}. This criterion states that local disorder is a relevant perturbation (in the renormalization group sense) when the correlation length critical exponent $\nu<2/d_s$, where $d_s$ is the spatial dimension. As recent work has investigated the effect disorder has on the measurement-induced phase transition \cite{ShkolnikZabaloVasseurHusePixleyGazit,ZabaloWilsonGullansVasseurGopalakrishnanHusePixley23}, investigating parallels between the effects of such disorder and potential counterparts in MBQC might also be potentially interesting.

\section{Acknowledgements}
\p This work was supported by the National Science Foundation STAQ Project (Grants No. PHY-1818914 and No. PHY-2325080) and Grant No. PHY-2310567. We thank the UNM Center for Advanced Research Computing, supported in part by the National Science Foundation, for providing the research computing resources used in this work.

\appendix
\section{Simulation Methods}
\label{sec:simmeth}
\subsection{Stabilizer formalism}
\label{subsec:stab}
\p A pitfall of using entanglement entropies to probe measurement-induced phase transitions is the drifting of the critical parameter $p_c$ for small system sizes \cite{ZabaloGullansWilsonGopalakrishnanHusePixley20}. Coupled with other finite size effects, it is often necessary to simulate large numbers of qubits to attain a quantitative hold on properties of the transition, especially for higher dimensions \cite{SierantSchiroLewensteinTurkeshi22}.
\p We reach these larger system sizes by employing stabilizer machinery \cite{Gottesman19,AaronsonGottesman04}. This well-known formalism tracks the quantum state $\rho$ in terms of its stabilizers $S(\rho)$, which are the group of operators of which the state is a $+1$ eigenstate. $S(\rho)$, which is exponentially large for pure states, is conveniently handled by looking at its generators. For the standard setting of qubits $S(\rho)$ is usually taken to be the Abelian subgroup of the Pauli group of tensor products of the single-qubit operators $\{I,X,Y,Z\}$, including an overall phase. In this case the generators themselves are Paulis with phases, and are computationally amiable. For this reason, Pauli stabilizer states are simply referred to as \textit{the} ``stabilizer states'' for short. 
\p The $n$-qubit Pauli operators constituting such a state's stabilizers may be represented (up to phase) by a string of $2n$ binary elements. By convention, the bits of such a string correspond to the presence or absence of an $X$ or $Z$ for a particular qubit, with a $Y$ understood by the presence of both. Often all $X$ bits are grouped together, followed by the $Z$ bits. Hence, the two-qubit operators $X_1X_2$, $Z_1X_2$, $I_1X_2$, and $Y_1Z_2$ may be represented by bitstrings $1100$, $0110$, $0100$, and $1011$, respectively. Thus, the $n$ generators of a phase-agnostic Pauli stabilizer state's stabilizer group may be represented by a $n\times 2n$ binary matrix. This technology results in considerable savings on space complexity compared to the naive representation of generic states.
\p One typically wishes not for some quantum state by itself, but rather that state as the input or output of some circuit consisting of unitary actions and measurements. Stabilizer states typically cease to be such under generic operations, so nonstabilizer states must be decomposed into a stabilizer basis, each component being simulated by itself and further splitting \cite{BrayviSmithSmolin16}. However, in the case where the circuits consist only of Pauli operators and a special subset of unitaries known as Clifford unitaries, stabilizer states are mapped to stabilizer states. Restriction of circuit elements to these operations permits efficient simulation of a state's evolution \cite{Gottesman19}.
\p Clifford unitaries are the normalizer $\mathcal{N}(\mathcal{P})$ of the set of Pauli products $\mathcal{P}$, i.e., the subset of unitaries $U\in\mathcal{N}(\mathcal{P})\subset U(2^n)$ such that $U^{-1}\mathcal{P}U=\mathcal{P}$. Limiting one's attention to Pauli products consisting only of one, two, three, \textit{etc}. nonidentity factors, one speaks of single-, two-, three-, \textit{etc}., qubit Clifford operators. When a Pauli group of symmetries is enforced on the system, the set of allowed Cliffords is further reduced by the requirement that they belong to the centralizer of that Pauli subgroup. The action of Clifford unitaries on a stabilizer state can be understood via their Heisenberg representation action on the state's stabilizer group. Representing that group's generators via binary matrices, Cliffords correspond to symplectic operators over $\mathbb{Z}_2$. Hence, not only are stabilizer states efficiently expressible, but so too are the natural unitary operations upon them.
\p We now turn to measurements. The measurement of an operator results in a projection of the state. As Pauli operators only have the two eigenvalues $\pm 1$, and since they square to the identity, this projection takes on the form $\Pi^{\pm}_P=\tfrac{1}{2}(\mathds{I}\pm P)$, where $P$ is the Pauli operator being measured. The density operator of a stabilizer state $\rho$ itself may itself be represented as a product of the projectors corresponding to the generators of its stabilizer group $S(\rho)$---this is a consequence of basic representation theory. We thus write the stabilizer state operator as

\eq{
    \rho=\prod\limits_{i=1}^n\Pi_{P_i}.
}

\p The action of $\Pi^{\pm}_P$ on the density matrix $\rho$ of a state can be understood by looking at its conjugate action on general projector $\Pi_{P'}$:

\eq{
    \Pi^{\pm}_{P}\Pi_{P'}\Pi^{\pm}_{P}&=\frac{\mathds{I}\pm P}{2}\frac{\mathds{I}+P'}{2}\frac{\mathds{I}\pm P}{2},\nn\\
    &=\frac{2(\mathds{I}\pm P)+(P'\pm(PP'+P'P)+PP'P)}{8}.
}

\p Two Paulis either commutate or anticommutate with one another. When they commutate, the above product becomes $\Pi_{P'}\Pi^{\pm}_{P}$. When they anticommutate, the result is $\tfrac{1}{2}\Pi^{\pm}_{P}$. As the stabilizer group is Abelian, the factors $\Pi_{P_i}$ of $\rho$ can be freely ordered. One may hence group all the $P_i$'s commutating with $P$ at the beginning, and move all those that anticommutate to the end. Let the generators commuting with $P$ be $\{R_i\}$ and those anticommutating be $\{Q_i\}$. Since any two $Q_i$'s commutate with each other, the identity $\Pi_{Q_i}\Pi_{Q_j}=\Pi_{Q_iQ_j}\Pi_{Q_i}$ is satisfied. The product of the anticommutating generators may be written

\eq{
    \prod\limits_{i=1}^m\Pi_{Q_i}&=\prod\limits_{i=1}^{m-1}\Pi_{Q_1Q_{i+1}}\Pi_{Q_1}.
}

\p The double products $Q_iQ_j$ commutate with $P$, and so may be absorbed into the commutating generators $R_i$. There is now only a single generator anticommutating with $P$. The action of $\Pi^{\pm}_{P}$ on $\rho$ can thus be written

\eq{
    \Pi^{\pm}_{P}\rho\Pi^{\pm}_{P}&=\Pi^{\pm}_{P}\left(\prod\limits_{i=1}^n\Pi_{P_i}\right)\Pi^{\pm}_{P},\nn\\
    &=\Pi^{\pm}_{P}\left(\prod\limits_i R_i\right)\Pi_{Q_1}\Pi^{\pm}_{P},\nn\\
    &=\left(\prod\limits_i R_i\right)\Pi^{\pm}_{P}\Pi_{Q_1}\Pi^{\pm}_{P},\nn\\
    &\to\left(\prod\limits_i R_i\right)\Pi&{\pm}_P,
}

where in the final line a factor of $\tfrac{1}{2}$ has been dropped after the state is renormalized.  The new list of stabilizers includes all the old commutating Paulis, all the old anticommutating Paulis (except $Q_1$) multiplied by $Q_1$, and the measured Pauli $P$ (replacing $Q_1$). A shorter explanation proceeds along the following logic. One first observes that $\rho$ may be represented using any choice of generators of $S(\rho)$. With this freedom, all anticommutating generators except for $Q_1$ can be replaced by $Q_iQ_1$. Then, in the expansion by projections one readily sees that applying $\Pi^{\pm}_P$ to the state has the effect of replacing the $Q_1$ projector with itself. As a whole, the algorithm can be described like so: if a measured operator $P$ anticommutates with some generators of the stabilize group of a state, replace one of the anticommutating generators $Q_1$ with $P$, then replace all other anticommutating $Q_i$ with $Q_1Q_i$.
\p We have ignored phases in this explanation, as they are not important for our purposes, however they introduce extra complications, even moreso when $P$ commutates with $S(\rho)$. Where the anticommutating case corresponds to a nondeterministic measurement, the commuting case represents a deterministic measurement. In this case, extra manipulations must be performed to determine the sign of the deterministic measurement. The ensuing complications may be ameliorated by more thoughtful implementations \cite{AaronsonGottesman04,Craig21}.
\p The stabilize formalism provides a twofold advantage: it significantly reduces space complexity, and it facilitates access to a class of highly entangled states. A general $n$-qubit wave function naively requires $\mathcal{O}(2^n)$ numbers, and quickly becomes unfeasible to work with. Tensor approximations, meanwhile, effectively model the low-entanglement area-law phase, but fail in the volume-law phase and near the critical point. Qubit stabilizers require only $\mathcal{O}(n^2)$ binaries, and are hardly less tractable in the volume-law phase than in the area. They therefore present an efficient probe of characteristics of measurement-induced phase transitions, provided one's purposes are met by the limited stabilizer polytope of Hilbert space we are constrained to.
\p Our specific code stores the binary string corresponding to a given Pauli in chunks, with each chunk being a 256-bit integer vector. On somewhat more modern architectures, these vectors may be increased in size. This vectorization makes various operations, such as the calculating the product of two Paulis, quite fast, particularly since our purposes are agnostic to phases. Further improvements could be made, such as implementing advanced rank-calculation algorithms for finite fields instead of employing binary Gaussian elimination \cite{BertolazziRimoldi14}.

\subsection{Unitary generation}
\label{subsec:generation}
    \p To simulate our random monitored circuits we must know how to implement the gates and measurements of interest. For measurements, this is simple. We only employ computational basis $Z$ measurements and 2D cluster state ``+''-shaped stabilizer measurements (see \refig{stencils}). Carrying out these measurements in the stabilizer formalism is particularly easy, as our purposes are unconcerned with stabilizer phase.
    \p For our three phase diagrams our gates are drawn from three corresponding unitary ensembles: unconstrained-, SPT-, and SSPT five-qubit Cliffords. This poses a predicament, as the set of all five-qubit Cliffords poses a prohibitive space requirement. The set of $n$-qubit Cliffords may be generated recursively given the set of $n-1$ qubit Cliffords via a well-known scheme \cite{LiChenFisher18}. At each step, one randomly selects an $n-1$ qubit Clifford and two anticommutating Pauli strings corresponding to what the $X$ and $Z$ operators map to for the newest qubit. Ignoring a phase, this amounts to $4^{k+1}-1$ possibilities for the first choice, and $4^k$ for the second. The number of unsigned $n$-qubit Cliffords may be obtained via induction:

    \eq{
        \text{\# }n\text{-qubit Cliffords}=2^{n^2}\prod_{i=1}^n \left(2^{2(i+1)}-1\right). \numberthis \label{eq:ccount}
    }

    \p This evaluates to $6, 720, 1451520, 47377612800,$ and\\ $24815256521932800$ for $n=1,2,3,4,5$ qubits, respectively. If each is naively represented by $2n\times 2n$ binary numbers, this corresponds to $24, 11520, 52254720, 3032167219200,$ and\\ $2481525652193280000$ bits, respectively, amounting to around $3b, 1\text{ KB}, 7\text{ MB}, 380 \text{ TB},$ and $300 \text{ PB}$ of space. Frontier currently boasts $700$ petabytes of storage capacity \cite{Frontier}, so this is not beyond the scope of modern computational prowess, but a more parsimonious usage of resources would be preferable
\p A satisfactory approach we have found generates the unconstrained Clifford and globally constrained SPT Clifford ensembles on the fly using the standard recursion algorithm. We store all three-qubit Cliffords, which may be generated very quickly. By randomly selecting anticommutating four-qubit and five-qubit Paulis, which have also been computed and stored in advance, we then generate random four-qubit and then five-qubit Cliffords. For the SPT Clifford ensemble, additional checks are performed to ensure that the resultant unitary is adequate---this process may be sped up by thoughtful implementation optimizations.
\p The SSPT Clifford ensemble is small enough to be collected, stored, and sampled from as needed. The set of all five-qubit Cliffords can be searched through by generating it in small enough chunks to be stored and have symmetry conditions checked. This procedure is readily parallelized. We have found that a better (also parallelizable) method starts by generating the small number of acceptable ``$Z$'' columns of such unitaries, then complete the remainder by generating and checking all compatible ``$X$'' columns. This last method may be realized in a matter of hours even with Python.

\section{Diagnostics}
\label{sec:diagnostics}
\begin{figure}
    \centering
    \includegraphics[width=0.85\linewidth]{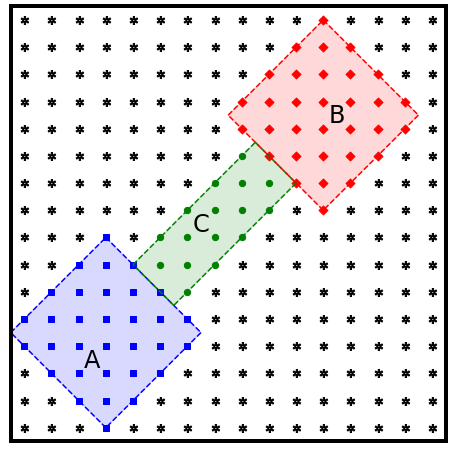}
    \caption{\label{fig:dumbbell} Example regions defined for the dumbbell entanglement entropy. In this work it is the tripartite mutual information of the heads conditioned on the handle, which is oriented along the direction of subsystem symmetries of the 2D cluster state.}
\end{figure}

\p In the conventional classification of ground states, SPT states are short-range entangled states from the strict topological perspective \cite{ChenGuLiuWen13}. This is to say that a shallow local circuit is able to transform them into product states with trivial entanglement. What defines symmetry-protected classes is that any such local circuit necessarily violates some natural symmetry. The requirement of this symmetry makes such a transition nontrivial, thus it is said that the symmetry eponymously protects the phase thus defined. It is usually not difficult to distinguish SPT states from volume-law states; most diagnostics distinguishing any area-law behavior from volume-law suffice. Distinguishing SPT area-law entanglement phases from trivial area-law phases, however, is more challenging. Furthermore, trying to identify strong SSPT states like the 2D cluster state---which is not a genuinely global SPT state (see Ref.~\cite{MillerMiyake16})---is an even more elusive task.
\p This section covers the three primary diagnostics we employ: the topological-, ancilla-, and dumbbell entanglement entropies. The first two consistently distinguish the cluster area-law phase from the volume-law phase, but do not distinguish between other area-law phases. The dumbbell entanglement entropy (see \refig{dumbbell}) successfully discriminates them, but is numerically less robust.

\subsection{Topological entanglement entropy}
\label{subsec:topologicalentanglement}
\begin{figure*}\customlabel{fig:unsclextraa}{16(a)}\customlabel{fig:unsclextrab}{16(b)}
    \centering
    \includegraphics[width=\linewidth]{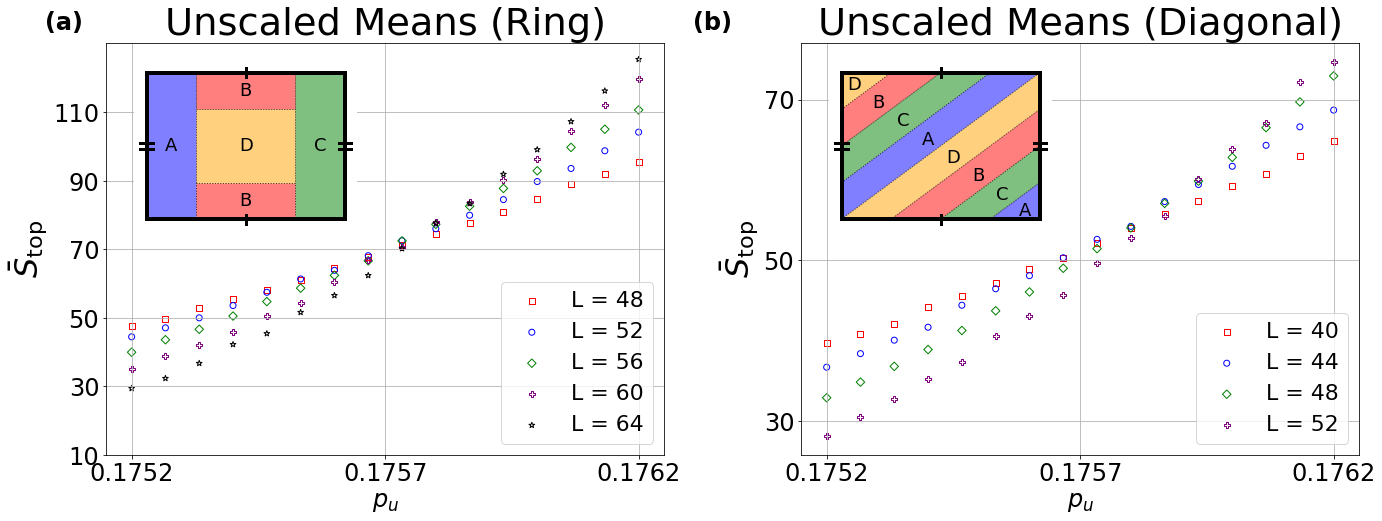}
    \caption{\label{fig:unsclextra} Unscaled data for seven-term circuit/trajectory-averaged topological entanglement entropy $\bar{S}_{\text{top}}$ for the SSPT Clifford ensemble with $p_M^z=0$ as a function of the probability of applying a unitary operation $p_u$ for the standard ``ring'' geometry (a) and ``diagonal'' geometry (b) oriented along the direction of subsystem symmetry (geometries inset). Analysis of this data is shown in Figs.~\ref{fig:sptvolmaxreg} and \ref{fig:sptvolmaxdiag}, respectively.}
\end{figure*}

\begin{figure*}[!ht]\customlabel{fig:subregace}{17a,c,e}\customlabel{fig:subregbdf}{17b,d,f}\customlabel{fig:subregab}{17a,b}\customlabel{fig:subregcd}{17c,d}\customlabel{fig:subregef}{17e,f}
    \centering
    \includegraphics[width=\linewidth]{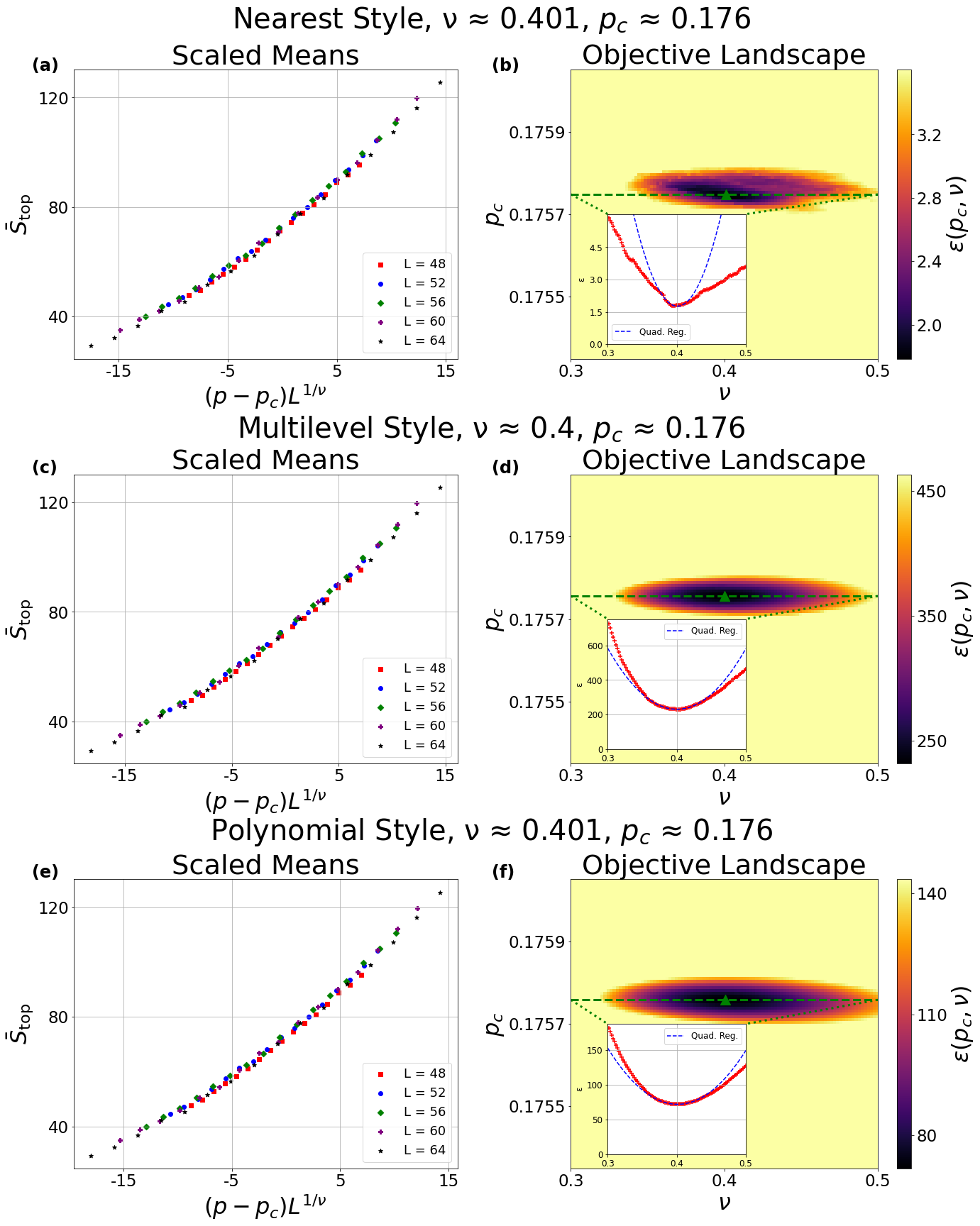}
    \caption{\label{fig:sptvolmaxreg} Analysis using the standard ``ring'' geometry. Data of the circuit/trajectory-averaged seven-term topological entanglement entropy $\bar{S}_{top}$ for the SSPT Clifford ensemble is depicted with $p_M^z=0$ for the cluster-to-volume transition. (a, b), (c, d), and (e, f) all illustrate collapse/objective landscapes (a, c, e/b, d, f) using three different fitting methods.}
\end{figure*}

\begin{figure*}[!ht]\customlabel{fig:subdiagace}{18a,c,e}\customlabel{fig:subdiagbdf}{18b,d,f}\customlabel{fig:subdiagab}{18a,b}\customlabel{fig:subdiagcd}{18c,d}\customlabel{fig:subdiagef}{18e,f}
    \centering
    \includegraphics[width=\linewidth]{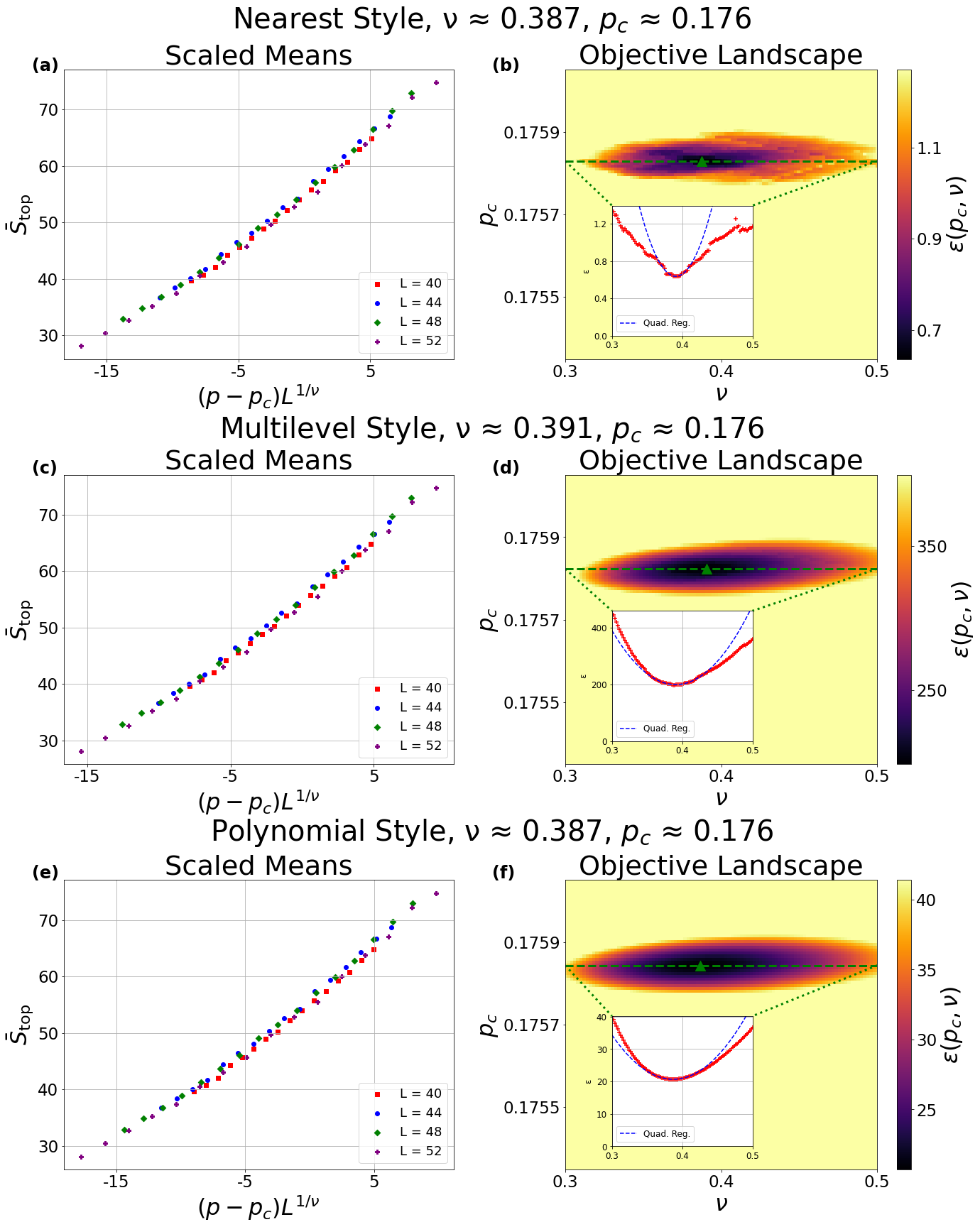}
    \caption{\label{fig:sptvolmaxdiag} Analysis using the ``diagonal'' geometry. Data of the circuit/trajectory-averaged seven-term topological entanglement entropy $\bar{S}_{top}$ for the SSPT Clifford ensemble is depicted with $p_M^z=0$ for the cluster-to-volume transition. (a, b), (c, d), and (e, f) all depict collapse (a, c, e) and objective landscapes (b, d, f) (with inset slices) using our three fitting methods.}
\end{figure*}

\p The ground states of a large class of gapped 2D systems generically feature area-law behavior. A subregion $R$'s entanglement entropy with respect the rest of the system follows $S_R\sim c|\partial R|-\Gamma$. Here $|\partial R|$ is the area-law contribution, while $\Gamma$ represents subleading corrections encoding long-range correlations. It is known that $\Gamma=\log\mathcal{D}$ for systems whose far-infrared behavior is well-described by topological quantum field theories, where $\mathcal{D}$ is the total quantum dimension of the effective theory. Extraction of $\Gamma$, which nicely characterizes the quantum phase in a single number, is typically carried out through clever combinations of entanglement entropies of varying regions; this is the topological entanglement entropy $S_{\text{top}}$ \cite{KitaevPreskill06,LevinWen06}.
\p The precise form comes in several varieties, possessing different numbers of terms and reflecting different geometries. The 2D cluster state has zero topological entanglement entropy for most conventional geometries, a property that is expected to be shared by a cluster area-law phase of our random monitored circuits. Nevertheless, variants of this quantity can still be used as an order parameter, since volume-law phase states will typically take on an extensive value. 
\p Our primary analysis uses the seven-term variant of $S_{\text{top}}$, which is related to the tripartite mutual information [Eq.~(\ref{eq:top})] \cite{ZengChenZhouWen19}. Our geometry, meanwhile, is defined in terms of cylinders embedded in a torus, as shown in \refig{q1dregions}. Other natural geometries (inset in \refig{unsclextra}) yield results quantitatively similar to those covered in the main text. A key result of our simulations is the appearance of a new universality class for the transition between the SSPT Clifford volume-law phase and the cluster area-law phase. As an example of this agreement, Figs.~\ref{fig:sptvolmaxreg} and \ref{fig:sptvolmaxdiag} both illustrate, for the SSPT Clifford ensemble along the $p_M^z=0.0$ line, the analysis and collapse of the circuit/trajectory-averaged mean $\bar{S}_{top}$ using standard ``ring'' and diagonally directed quasi-1D geometries. The critical point is, as for \refig{sptvolmaxq1d} in the main text, found to be $p_c\approx 0.176$ uniformly, while the critical exponent $\nu$ varies from $0.387$ to $0.406$.
\p Our sampling method is as follows. We first wait for a given instance of the circuit to reach a sort of steady state. This happens when the entanglement growth plateaus, which happens before some fraction of $L^4$ (we take that fraction to be $1/4$ for concreteness). Then, we sample $S_{\text{top}}$ every $L^2$ timesteps, until time $\mathcal{O}(L^4$), at which point we start an entirely new circuit again from scratch with a fresh product state input. The sampling frequency is safely uncorrelated far away from the critical point, but as the critical point is approached the correlation time formally diverges. Nevertheless, we have found that the resultant statistical compromise has an effect negligible compared to the effects of the accompanying growth of the statistical dispersion of the order parameter distribution of $S_{\text{top}}$. Our typical number of sample circuits depends on the system size, but varies from thousands to tends of thousands for smaller and larger systems, respectively.

\subsection{Ancilla entanglement entropy}
\label{subsec:ancilla}
\p One element underlying the stability of the volume-law phase of random monitored circuits is the notion of scrambling. Provided that the measurement rate is not too great, information about the initial state is, after some initial mixing time, encoded in long-range correlations that local measurements cannot efficiently probe. As a diagnostic, the ancilla entanglement entropy $\bar{S}_{\text{anc}}$ takes advantage of this phenomenon \cite{GullansHuse20p,GullansHuse20}. It is a measure of how much information about a probe entangled with the system (size $N$) remains encoded at a time $t$. Quantitatively, the unaveraged $S_{\text{anc}}$ is the von Neumann entropy of a probe entangled with the system:

\eq{
    S_{\text{anc}}&=S_{vN}(\rho_{anc}),
}
where $\rho_{\text{anc}}$ is the state operator of the ancillary system. As with most probes, this must be evaluated for an individual trajectory, then averaged over measurement outcomes and circuit realizations to yield the diagnostic $\bar{S}_{\text{anc}}$.
\p So long as a few physical qubits have not been measured some correlation will persist between the system and the ancilla qubits. Consequently, the von Neumann entanglement will be nonzero. Deep in an area-law phase, where measurements are frequent, the survival time is expected to scale polynomially in the system size. In contrast, scrambling protects information exponentially long in the volume-law phase. These timescales are not to be confused with the behavior of the decay itself, which is reasonably matched by an exponential fit after early times. The upper panels of \refig{telanc}, for example, show how the long-time behavior of $\bar{S}_{\text{anc}}$ exhibits exponential decay in the area-law phase near criticality. It is the survival time---not the behavior with respect to time---of this decay that rapidly increases in the volume-law phase and decreases in the area-law phase. For example, it is the $\tau$ in $e^{-t/\tau}$. Thus, by evaluating the entanglement at a time $t$ polynomial in the system size, we can expend a significant fraction of the survival time for ancilla qubits in an area-law system, while still keeping it safely scrambled for volume-law cases. This distinguishing behavior becomes more prominent as system size is increased, so $\bar{S}_{\text{anc}}$ can be effectively used to distinguish area-law phase behavior from volume-law.
\p Our implementation is as follows. Before starting the circuit properly, we entangle $20$ ancilla qubits by running a scrambling circuit for time $T=\mathcal{O}(L^4)$, with the five-qubit Clifford unitaries enacted on any arbitrary five qubits (either ancillary or system), regardless of spatial separation. Then, we sample the entanglement between the ancilla qubits and the system after $\mathcal{O}(L^4)$ timesteps. Immediately after scrambling, the entanglement is on order of $20$, and deep in the volume-law phase this value is retained for long periods of time. In the area-law phases, meanwhile, the entanglement drops quickly to zero. At criticality, a constant fraction of the original entanglement is retained in the thermodynamic limit, with the specific factor depending on when the entanglement is sampled. 
\p Because of the initial mixing time, and since $\bar{S}_{\text{anc}}$ is only really sampled once per circuit run, analysis of $\bar{S}_{\text{anc}}$ requires more simulation time. It makes up for this defect by being formally less noisy in spite of this defect. Typical sample sizes ranges from high hundreds for the largest systems to high thousands for smaller ones.

\subsection{Dumbbell entanglement entropy}
\label{subsec:dumbbell}
\p While the topological entanglement entropy succeeds in distilling the topological essence of many nice models down to a single number, so-called ``spurious'' contributions can arise in systems exhibiting long-range string order \cite{WilliamsonDuaCheng19,StephenDreyerIqbalSchuch19}. Though this fact spoils elegant theoretical models, its existence can be used as a diagnostic of the presence of SPT and SSPT effects. The dumbbell entanglement entropy $\bar{S}_{\text{dumb}}$ \cite{WilliamsonDuaCheng19} is a construction taking advantage of this fact. For this work we define it as the trajectory average of a seven-term tripartite mutual information as in Eq.~(\ref{eq:top}), rather than the conditional mutual information of \cite{WilliamsonDuaCheng19}. As opposed to the more standard geometries used for $\bar{S}_{top}$, the dumbbell entanglement entropy $\bar{S}_{\text{dumb}}$ is so named for using regions forming dumbbell with the handle directed along a line of string symmetry (see \refig{dumbbell}).
\p We make use of this quantity in this work to distinguish the cluster area-law phase, exhibiting subsystem symmetry, from the trivial area-law phase generated by an excess of local computational basis measurements. While it succeeds in discriminating between the two behaviors, in practice it is less robust at its job than the topological- or ancilla entanglement entropies covered above. The defects in the curve (seen in \refig{puremeasurement}) become more prominent near criticality, and capriciously depend on the size of the ``head'' and ``handle.''
\p $S_{\text{dumb}}$ is sampled similarly to $S_{\text{top}}$ as described in Sec.~\ref{subsec:topologicalentanglement}. We begin sampling at $t=\tfrac{1}{4}L^2$, and sample every $L^2$ timesteps, starting another circuit from scratch at $t=L^4$. This sampling frequency produces data sufficiently uncorrelated away from the critical point, but near the critical point the correlation time diverges along with the correlation length. Nevertheless, the statistical influence of this fact appears negligible compared to the growth of the spread of the order parameter distribution of $S_{\text{dumb}}$, whose circuit/trajectory mean $\bar{S}_{\text{dumb}}$ is our reported diagnostic. Typical sample sizes range from thousands to tens of thousands for larger and smaller systems, respectively.

\section{Fitting Methods}
\label{sec:fit}
\p Although the data presented in the main text is analyzed using a single fitting method, we apply three different techniques as a check on our finite-size scaling analysis. We refer to them as ``Nearest,'' ``Multilevel,'' and ``Polynomial'' styles, respectively. All take as input the critical parameters $(p_c,\nu,\gamma)$, which scale the data as $(p,\Delta)\to((p-p_c)L^{1/\nu},\Delta L^{-\gamma})$. Although we have included the extensive exponent $\gamma$ here, we have found that this is within margin of error $0$ in practice for all diagnostics of interest. It is therefore set to be exactly $0$ in all figures in this work.
\p For the ``Nearest'' style, we order the scaled data following $|p-p_c|L^{1/\nu}$. A linear regression is performed for all interior points using their immediate neighbors, regardless of system size. That is, for three datapoints $(x_{i-1}=(p_j-p_c)L_j^{\tfrac{1}{\nu}},y_{i-1}),(x_{i}=(p_{j'}-p_c)L_{j'}^{\tfrac{1}{\nu}},y_{i}),$ and $(x_{i+1}=(p_{j''}-p_c)L_{j''}^{\tfrac{1}{\nu}},y_{i+1})$ with $x_{i-1}\le x_i\le x_{i+1}$ we write for the regression
\eq{
    \tilde{y}_i=y_{i-1}+\tfrac{y_{i+1}-y_{i-1}}{x_{i+1}-x_{i-1}}(x_i-x_{i-1}),
}
\p The objective function $\epsilon(p_c,\nu)$ of this method is the average sum of residuals squared for each individual regression,
\eq{
    \epsilon(p_c,\nu)=\frac{1}{\vert\text{regress pts.}\vert}\sum\limits_{i} \left(y-\tilde{y}\left((p_i-p_c)L_i^{\tfrac{1}{\nu}}\right)\right)^2,
}
where $y$ is the diagnostic of interest, $\tilde{y}$ is the regression, $i$ indexes the data points, and $p_i$ and $L_i$ are the tuning parameter and system length associated with datapoint $i$. This method heuristically has the effect of favoring input parameters producing a curve locally as smooth as possible. Notably, this method does not make use of any estimated error of $y$, since each regression only uses two points.
\p Similar optimization procedures are carried out for our other methods as well. The second technique, ``Multilevel,'' looks for each individual point at every system size greater than that of the point, and performs a regression using the nearest neighbors from each level. The objective function is the sum of the regression deviation squared for each point, weighed by the estimated error. More specifically, for a given $(x_{i}=(p_j-p_c)L_j^{\tfrac{1}{\nu}},y_{i})$, we consider for each $L_{j'}>L_j$ the collection of nearest points $(x_{i'}=(p_j'-p_c)L_{j'}^{\tfrac{1}{\nu}},y_{i'})$, then use them in a linear regression to obtain $\tilde{y}\left((p_j-p_c)L_j^{\tfrac{1}{\nu}}\right)$. The objective function $\epsilon(p_c,\nu)$ is the sum of the deviation squared for each point based on this regression
\eq{
    \epsilon(p_c,\nu)=\frac{1}{\text{\# of points}}\sum\limits_{i} \left(\frac{y-\tilde{y}\left((p_i-p_c)L_i^{\tfrac{1}{\nu}}\right)}{\sigma_i}\right)^2,
}
where $\sigma_i$ is the estimated error for $y$, which can be of use here since each regression uses more than two points. The intuition underlying this method is that it favors values of $(p_c,\nu)$ specifying a curve as smooth as possible with respect to the samples coming from larger systems, which are less susceptible to finite size effects.
\p The final ``Polynomial'' style is a simple polynomial regression on the entirety of the scaled data, weighed by the estimated errors $\sigma_i$. $\epsilon(p_c,\nu)$ is simply the weighted square deviation of this regression:
\eq{
    \epsilon(p_c,\nu)=\sum\limits_{i} \left(\frac{y-\tilde{y}\left((p_i-p_c)L_i^{\tfrac{1}{\nu}}\right)}{\sigma_i}\right)^2.
}
\p As there is no averaging over the number of points, $\epsilon$ in this case could be used as a $\chi^2$ statistic, or for maximum likelihood estimation, but in practice this is erroneous, as the order parameter distributions feature heavy tails. nevertheless, the output of this method, crude as it is, produces smooth landscapes whose minima do not deviate significantly from the prior two methods, provided that the data is well-sampled and smooth. Eighth order polynomials have proven sufficient for our purposes; standard hypothesis testing techniques show us that higher orders suffer from overfitting, while lower orders fail to capture the common curve's contour.  This final method takes all points into account at once, and has the virtue of producing the smoothest objective landscapes with the most readily interpretable error margins. For this reason all figures in the main text employ the ``polynomial'' method. However, as shown in Figs. \ref{fig:sptvolmaxreg} and \ref{fig:sptvolmaxdiag}, all three methods are typically in reasonable agreement.
\p For each objective function we first guess $p_c$ and $\nu$, then run the curve-fitting routine and optimize with respect to $p_c$ and $\nu$ using the Nelder-Mead algorithm. Near the minimum the behavior of $\epsilon$ may be approximated by a regression; our error bars in Table~\ref{tab:exp} arise from this, corresponding roughly to the width of twice the minimum of $\epsilon(p_c,\nu)$. It is worth keeping in mind that this is not a statistically principled error bar reflecting confidence or credibility intervals, but rather a heuristic concerning the goodness of fit of different parameters. The error bars on the raw data points in our figures, meanwhile, lie within the marker size.

\section{Behavior at Pure Measurement Critical Point}
\label{sec:critmeas}
\begin{figure*}[!htbp]\customlabel{fig:subentgrowtha}{19(a)}\customlabel{fig:subentgrowthb}{19(b)}\customlabel{fig:subentgrowthc}{19(c)}
    \centering
    \includegraphics[width=0.8\linewidth]{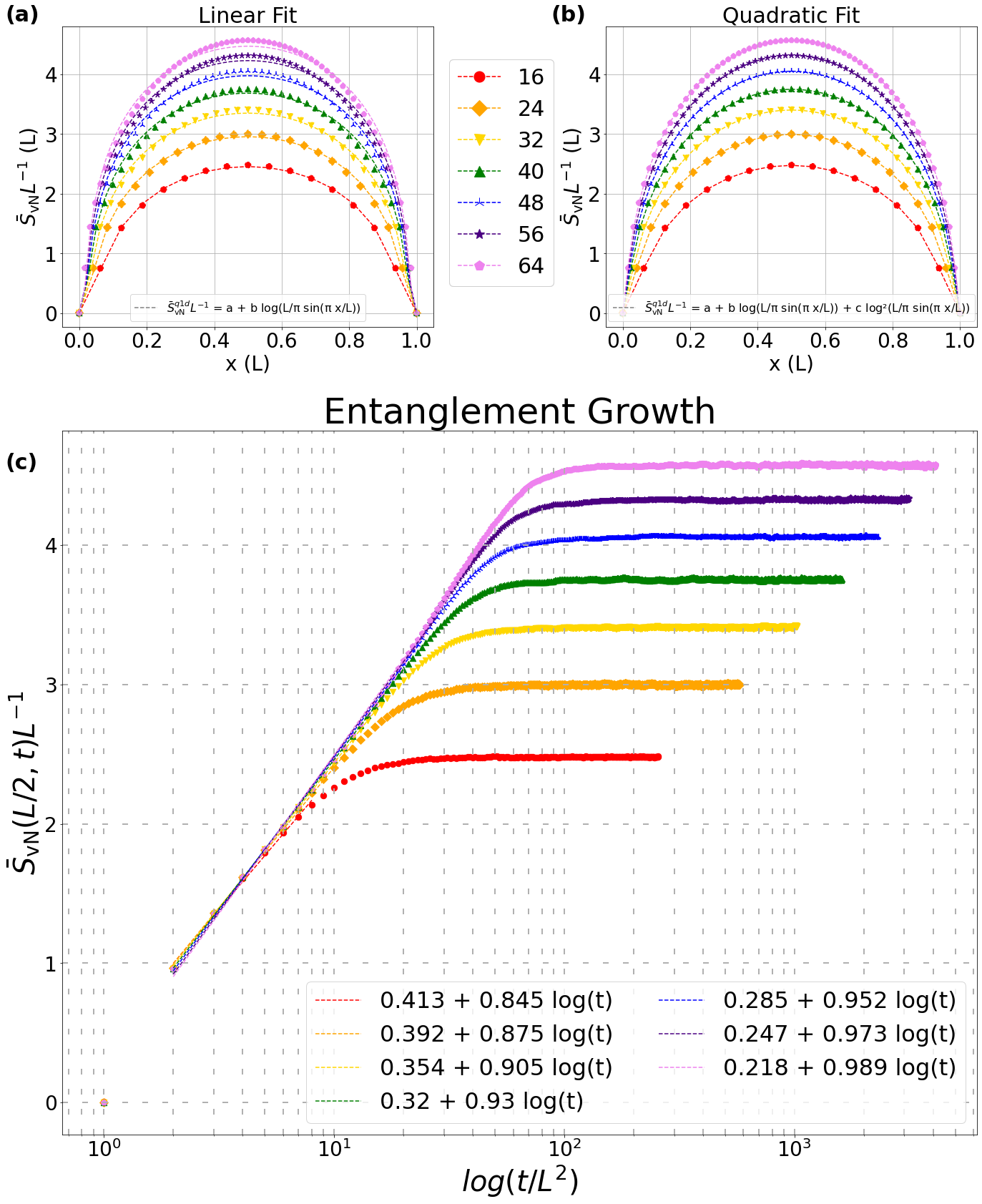}
    \caption{\label{fig:entropygrowth} Plots of the scaled entanglement versus system length for linear (a) and quadratic (b) regressions, and half-system entanglement growth versus time (c). Colors/shapes refer to the same system sizes across all figures. The linear fit works well for small $L$, but consistently deviates for larger sizes, while the quadratic fit continues well. Meanwhile, the regression coefficients for the growth over time slowly vary with $L$.}
\end{figure*}
\p For $(1+1)$D systems at criticality it is well-known that the entanglement entropy of a contiguous section (size $l$) of the system (size $L$) often features entanglement entropy scaling as

\eq{
    S_{\text{1D}}(l)=\text{Const.}+\tfrac{c}{6}\log\left(\tfrac{L}{\pi}\sin(\tfrac{\pi l}{L})\right), \numberthis\label{eq:sclprof}
}
where $c$ is the central charge of the associated conformal field theory \cite{CalabreseCardy04,CalabreseCardy09}. The entanglement entropy of a region of set size (say, half the system), meanwhile, grows logarithmically in time $S_{\text{1D}}(t)=a+b\log(t)$.
\p In 2D translationally invariant systems it is expected that spiritually similar expressions continue to hold, provided that quasi-1D sections of the system (such as cylindrical sections of a torus) are chosen. Furthermore, while the critical properties of random monitored quantum circuits do not precisely match the phenomenology of more conventional phase transitions, similar relationships still hold in the $(1+1)$D setting with certain modifications. For example, the coefficient of proportionality in $S_{\text{1D}}$ above is replaced by the primary weight of a boundary changing operator \cite{LiChenLudwigFisher21}. 
\p Between these principles it is reasonable to check the hypothesis of this scaling for our $(2+1)$D system. Other $(2+1)$D monitored circuits have found reasonable agreement for certain transitions \cite{LavasaniAlaviradBarkeshli21p}. We find that the agreement for the transition occurring in pure measurement dynamics [shown in \refig{subentgrowtha}] is respectable. However, for larger system sizes disagreement begins to emerge. Adding a term quadratic in $\log\left(\tfrac{L}{\pi}\sin(\tfrac{\pi x}{L})\right)$ appreciably resolves this discrepancy for the system sizes considered, as shown in \refig{subentgrowthb}. Quantitatively, the order of the square residual is two orders of magnitude smaller for this quadratic regression. Although we have no theoretical justification for this addition, it suggests that some power series in this variable captures the finer critical features observed. We also find in \refig{subentgrowthc} that the terms in the logarithmic linear regression for $S_{\text{1D}}(t)$ vary with system size as well.

\bibliography{ms}
\end{document}